\documentclass[longauth]{aaEC}

\usepackage{graphicx}
\usepackage{natbib}
\usepackage{scalerel}
\usepackage{siunitx}
\usepackage{lastpage}
\usepackage{soul}
\usepackage{amsmath}
\newcommand{\linenumbers}[0]{}

\usepackage[switch, modulo]{lineno}

\usepackage[table]{xcolor}

\newcolumntype{P}[1]{>{\centering\arraybackslash}p{#1}}
\newcommand*{\rom}[1]{\expandafter\@slowromancap\romannumeral #1@}
\newcommand{\orcid}[1]{}

\usepackage{txfonts}
\usepackage[pdfencoding=auto,psdextra]{hyperref}
\hypersetup{
    colorlinks=true,
    linkcolor=blue,
    filecolor=magenta,      
    urlcolor=blue,
    citecolor=blue
}
\makeatletter
\renewcommand*\aa@pageof{, page \thepage{} of \pageref*{LastPage}}
\makeatother

\usepackage[utf8]{inputenc}
\usepackage{datetime2}

\renewcommand{\linenumbers}[0]{}

\usepackage{euclid}

\begin{document}

\title{\Euclid: Disky titans -- surprisingly high star formation activity and gas content in two brightest group galaxies at z$\sim$0.75\thanks{This paper is published on behalf of the Euclid Consortium.}}

%%%% Version Monday 28th of September 2026 11:55:51 AM UT
%%%% Assumes the new A&A style file from Oct 2025 or later
%%%% Please do not edit the author list -- contact ECEB Bureau for changes
\author{F.~Gentile\orcid{0000-0002-8008-9871}\thanks{\email{fabrizio.gentile@cea.fr}}\inst{\ref{aff1},\ref{aff2}}
\and E.~Daddi\orcid{0000-0002-3331-9590}\inst{\ref{aff3}}
\and D.~Elbaz\orcid{0000-0002-7631-647X}\inst{\ref{aff3}}
\and A.~Enia\orcid{0000-0002-0200-2857}\inst{\ref{aff2}}
\and F.~Vito\orcid{0000-0003-0680-9305}\inst{\ref{aff2}}
\and P.-A.~Duc\orcid{0000-0003-3343-6284}\inst{\ref{aff4}}
\and M.~Franco\orcid{0000-0002-3560-8599}\inst{\ref{aff3}}
\and H.~Fu\orcid{0009-0002-8051-1056}\inst{\ref{aff5},\ref{aff6}}
\and R.~Giuffrida\orcid{0000-0002-2774-3491}\inst{\ref{aff3}}
\and D.~Roberts\orcid{0009-0009-7662-0445}\inst{\ref{aff6}}
\and F.~Shankar\orcid{0000-0001-8973-5051}\inst{\ref{aff6}}
\and S.~Lu\orcid{0000-0001-5988-2202}\inst{\ref{aff7},\ref{aff8},\ref{aff9}}
\and P.~Awad\orcid{0000-0002-0428-849X}\inst{\ref{aff10}}
\and J-B.~Billand\orcid{0009-0004-4168-3634}\inst{\ref{aff1}}
\and M.~Baes\orcid{0000-0002-3930-2757}\inst{\ref{aff11}}
\and L.~Bisigello\orcid{0000-0003-0492-4924}\inst{\ref{aff12}}
\and E.~Duran-Camacho\orcid{0000-0002-3153-0536}\inst{\ref{aff13},\ref{aff14}}
\and G.~Castignani\orcid{0000-0001-6831-0687}\inst{\ref{aff2}}
\and O.~Cucciati\orcid{0000-0002-9336-7551}\inst{\ref{aff2}}
\and G.~De~Lucia\orcid{0000-0002-6220-9104}\inst{\ref{aff15}}
\and C.~D'Eugenio\orcid{0000-0001-7344-3126}\inst{\ref{aff16},\ref{aff1}}
\and D.~Donevski\orcid{0000-0001-5341-2162}\inst{\ref{aff17},\ref{aff18}}
\and M.~Fossati\orcid{0000-0002-9043-8764}\inst{\ref{aff19},\ref{aff20}}
\and M.~Fumagalli\orcid{0000-0001-6676-3842}\inst{\ref{aff19},\ref{aff15}}
\and R.~Gobat\orcid{0000-0002-9566-1922}\inst{\ref{aff21}}
\and C.~Gruppioni\orcid{0000-0002-5836-4056}\inst{\ref{aff2}}
\and M.~Magliocchetti\orcid{0000-0001-9158-4838}\inst{\ref{aff22}}
\and B.~Magnelli\orcid{0000-0002-6777-6490}\inst{\ref{aff3}}
\and G.~Papini\orcid{0000-0002-5810-318X}\inst{\ref{aff2},\ref{aff23}}
\and L.~Pozzetti\orcid{0000-0001-7085-0412}\inst{\ref{aff2}}
\and V.~Sangalli\orcid{0009-0006-9595-4393}\inst{\ref{aff3}}
\and J.~G.~Sorce\orcid{0000-0002-2307-2432}\inst{\ref{aff24},\ref{aff25}}
\and L.~Spinoglio\orcid{0000-0001-8840-1551}\inst{\ref{aff22}}
\and V.~Strazzullo\inst{\ref{aff15},\ref{aff26}}
\and M.~Tarrasse\orcid{0009-0009-3123-4479}\inst{\ref{aff1}}
\and G.~Toni\orcid{0009-0005-3133-1157}\inst{\ref{aff23},\ref{aff2},\ref{aff27}}
\and G.~Zamorani\orcid{0000-0002-2318-301X}\inst{\ref{aff2}}
\and J.~Lewis\orcid{0000-0001-7917-8474}\inst{\ref{aff1}}
\and B.~Altieri\orcid{0000-0003-3936-0284}\inst{\ref{aff28}}
\and S.~Andreon\orcid{0000-0002-2041-8784}\inst{\ref{aff20}}
\and N.~Auricchio\orcid{0000-0003-4444-8651}\inst{\ref{aff2}}
\and C.~Baccigalupi\orcid{0000-0002-8211-1630}\inst{\ref{aff26},\ref{aff15},\ref{aff29},\ref{aff17}}
\and M.~Baldi\orcid{0000-0003-4145-1943}\inst{\ref{aff30},\ref{aff2},\ref{aff31}}
\and S.~Bardelli\orcid{0000-0002-8900-0298}\inst{\ref{aff2}}
\and P.~Battaglia\orcid{0000-0002-7337-5909}\inst{\ref{aff2}}
\and A.~Biviano\orcid{0000-0002-0857-0732}\inst{\ref{aff15},\ref{aff26}}
\and M.~Bolzonella\orcid{0000-0003-3278-4607}\inst{\ref{aff2}}
\and E.~Branchini\orcid{0000-0002-0808-6908}\inst{\ref{aff32},\ref{aff33},\ref{aff20}}
\and M.~Brescia\orcid{0000-0001-9506-5680}\inst{\ref{aff34},\ref{aff35}}
\and S.~Camera\orcid{0000-0003-3399-3574}\inst{\ref{aff36},\ref{aff37},\ref{aff38}}
\and V.~Capobianco\orcid{0000-0002-3309-7692}\inst{\ref{aff38}}
\and C.~Carbone\orcid{0000-0003-0125-3563}\inst{\ref{aff39}}
\and J.~Carretero\orcid{0000-0002-3130-0204}\inst{\ref{aff40},\ref{aff41}}
\and M.~Castellano\orcid{0000-0001-9875-8263}\inst{\ref{aff42}}
\and S.~Cavuoti\orcid{0000-0002-3787-4196}\inst{\ref{aff35},\ref{aff43}}
\and A.~Cimatti\inst{\ref{aff44}}
\and C.~Colodro-Conde\inst{\ref{aff13}}
\and G.~Congedo\orcid{0000-0003-2508-0046}\inst{\ref{aff45}}
\and L.~Conversi\orcid{0000-0002-6710-8476}\inst{\ref{aff46},\ref{aff28}}
\and Y.~Copin\orcid{0000-0002-5317-7518}\inst{\ref{aff47}}
\and F.~Courbin\orcid{0000-0003-0758-6510}\inst{\ref{aff48},\ref{aff49},\ref{aff50}}
\and H.~M.~Courtois\orcid{0000-0003-0509-1776}\inst{\ref{aff51}}
\and M.~Cropper\orcid{0000-0003-4571-9468}\inst{\ref{aff52}}
\and H.~Degaudenzi\orcid{0000-0002-5887-6799}\inst{\ref{aff53}}
\and H.~Dole\orcid{0000-0002-9767-3839}\inst{\ref{aff25}}
\and F.~Dubath\orcid{0000-0002-6533-2810}\inst{\ref{aff53}}
\and X.~Dupac\inst{\ref{aff28}}
\and M.~Farina\orcid{0000-0002-3089-7846}\inst{\ref{aff22}}
\and R.~Farinelli\inst{\ref{aff2}}
\and F.~Faustini\orcid{0000-0001-6274-5145}\inst{\ref{aff42},\ref{aff54}}
\and S.~Ferriol\inst{\ref{aff47}}
\and S.~Fotopoulou\orcid{0000-0002-9686-254X}\inst{\ref{aff55}}
\and M.~Frailis\orcid{0000-0002-7400-2135}\inst{\ref{aff15}}
\and E.~Franceschi\orcid{0000-0002-0585-6591}\inst{\ref{aff2}}
\and M.~Fumana\orcid{0000-0001-6787-5950}\inst{\ref{aff39}}
\and S.~Galeotta\orcid{0000-0002-3748-5115}\inst{\ref{aff15}}
\and K.~George\orcid{0000-0002-1734-8455}\inst{\ref{aff56}}
\and B.~Gillis\orcid{0000-0002-4478-1270}\inst{\ref{aff45}}
\and C.~Giocoli\orcid{0000-0002-9590-7961}\inst{\ref{aff2},\ref{aff31}}
\and J.~Gracia-Carpio\orcid{0000-0003-4689-3134}\inst{\ref{aff57}}
\and A.~Grazian\orcid{0000-0002-5688-0663}\inst{\ref{aff12}}
\and F.~Grupp\inst{\ref{aff57},\ref{aff58}}
\and S.~Gwyn\orcid{0000-0001-8221-8406}\inst{\ref{aff59}}
\and S.~V.~H.~Haugan\orcid{0000-0001-9648-7260}\inst{\ref{aff60}}
\and H.~Hoekstra\orcid{0000-0002-0641-3231}\inst{\ref{aff10}}
\and W.~Holmes\orcid{0009-0007-8554-4646}\inst{\ref{aff61}}
\and I.~M.~Hook\orcid{0000-0002-2960-978X}\inst{\ref{aff62}}
\and F.~Hormuth\inst{\ref{aff63}}
\and A.~Hornstrup\orcid{0000-0002-3363-0936}\inst{\ref{aff64},\ref{aff65}}
\and M.~Huertas-Company\orcid{0000-0002-1416-8483}\inst{\ref{aff13},\ref{aff66},\ref{aff67}}
\and K.~Jahnke\orcid{0000-0003-3804-2137}\inst{\ref{aff68}}
\and M.~Jhabvala\inst{\ref{aff69}}
\and B.~Joachimi\orcid{0000-0001-7494-1303}\inst{\ref{aff70}}
\and S.~Kermiche\orcid{0000-0002-0302-5735}\inst{\ref{aff71}}
\and A.~Kiessling\orcid{0000-0002-2590-1273}\inst{\ref{aff61}}
\and B.~Kubik\orcid{0009-0006-5823-4880}\inst{\ref{aff47}}
\and M.~K\"ummel\orcid{0000-0003-2791-2117}\inst{\ref{aff58}}
\and M.~Kunz\orcid{0000-0002-3052-7394}\inst{\ref{aff72}}
\and H.~Kurki-Suonio\orcid{0000-0002-4618-3063}\inst{\ref{aff73},\ref{aff74}}
\and A.~M.~C.~Le~Brun\orcid{0000-0002-0936-4594}\inst{\ref{aff75}}
\and S.~Ligori\orcid{0000-0003-4172-4606}\inst{\ref{aff38}}
\and P.~B.~Lilje\orcid{0000-0003-4324-7794}\inst{\ref{aff60}}
\and V.~Lindholm\orcid{0000-0003-2317-5471}\inst{\ref{aff73},\ref{aff74}}
\and I.~Lloro\orcid{0000-0001-5966-1434}\inst{\ref{aff76}}
\and G.~Mainetti\orcid{0000-0003-2384-2377}\inst{\ref{aff77}}
\and O.~Mansutti\orcid{0000-0001-5758-4658}\inst{\ref{aff15}}
\and O.~Marggraf\orcid{0000-0001-7242-3852}\inst{\ref{aff78}}
\and M.~Martinelli\orcid{0000-0002-6943-7732}\inst{\ref{aff42},\ref{aff79}}
\and N.~Martinet\orcid{0000-0003-2786-7790}\inst{\ref{aff80}}
\and F.~Marulli\orcid{0000-0002-8850-0303}\inst{\ref{aff23},\ref{aff2},\ref{aff31}}
\and R.~J.~Massey\orcid{0000-0002-6085-3780}\inst{\ref{aff81}}
\and E.~Medinaceli\orcid{0000-0002-4040-7783}\inst{\ref{aff2}}
\and S.~Mei\orcid{0000-0002-2849-559X}\inst{\ref{aff82},\ref{aff83}}
\and M.~Melchior\inst{\ref{aff84}}
\and M.~Meneghetti\orcid{0000-0003-1225-7084}\inst{\ref{aff2},\ref{aff31}}
\and E.~Merlin\orcid{0000-0001-6870-8900}\inst{\ref{aff42}}
\and G.~Meylan\orcid{0000-0001-6503-0209}\inst{\ref{aff85}}
\and A.~Mora\orcid{0000-0002-1922-8529}\inst{\ref{aff86}}
\and M.~Moresco\orcid{0000-0002-7616-7136}\inst{\ref{aff23},\ref{aff2}}
\and L.~Moscardini\orcid{0000-0002-3473-6716}\inst{\ref{aff23},\ref{aff2},\ref{aff31}}
\and R.~Nakajima\orcid{0009-0009-1213-7040}\inst{\ref{aff78}}
\and C.~Neissner\orcid{0000-0001-8524-4968}\inst{\ref{aff87},\ref{aff41}}
\and S.-M.~Niemi\orcid{0009-0005-0247-0086}\inst{\ref{aff88}}
\and C.~Padilla\orcid{0000-0001-7951-0166}\inst{\ref{aff87}}
\and S.~Paltani\orcid{0000-0002-8108-9179}\inst{\ref{aff53}}
\and F.~Pasian\orcid{0000-0002-4869-3227}\inst{\ref{aff15}}
\and K.~Pedersen\inst{\ref{aff89}}
\and W.~J.~Percival\orcid{0000-0002-0644-5727}\inst{\ref{aff90},\ref{aff91},\ref{aff92}}
\and V.~Pettorino\orcid{0000-0002-4203-9320}\inst{\ref{aff88}}
\and G.~Polenta\orcid{0000-0003-4067-9196}\inst{\ref{aff54}}
\and M.~Poncet\inst{\ref{aff93}}
\and L.~A.~Popa\inst{\ref{aff94}}
\and F.~Raison\orcid{0000-0002-7819-6918}\inst{\ref{aff57}}
\and A.~Renzi\orcid{0000-0001-9856-1970}\inst{\ref{aff95},\ref{aff96},\ref{aff2}}
\and J.~Rhodes\orcid{0000-0002-4485-8549}\inst{\ref{aff61}}
\and G.~Riccio\inst{\ref{aff35}}
\and E.~Romelli\orcid{0000-0003-3069-9222}\inst{\ref{aff15}}
\and M.~Roncarelli\orcid{0000-0001-9587-7822}\inst{\ref{aff2}}
\and H.~J.~A.~Rottgering\orcid{0000-0001-8887-2257}\inst{\ref{aff10}}
\and B.~Rusholme\orcid{0000-0001-7648-4142}\inst{\ref{aff97}}
\and R.~Saglia\orcid{0000-0003-0378-7032}\inst{\ref{aff58},\ref{aff57}}
\and Z.~Sakr\orcid{0000-0002-4823-3757}\inst{\ref{aff98},\ref{aff99},\ref{aff100}}
\and D.~Sapone\orcid{0000-0001-7089-4503}\inst{\ref{aff101}}
\and P.~Schneider\orcid{0000-0001-8561-2679}\inst{\ref{aff78}}
\and T.~Schrabback\orcid{0000-0002-6987-7834}\inst{\ref{aff102}}
\and A.~Secroun\orcid{0000-0003-0505-3710}\inst{\ref{aff71}}
\and E.~Sihvola\orcid{0000-0003-1804-7715}\inst{\ref{aff103}}
\and P.~Simon\inst{\ref{aff78}}
\and C.~Sirignano\orcid{0000-0002-0995-7146}\inst{\ref{aff95},\ref{aff96}}
\and G.~Sirri\orcid{0000-0003-2626-2853}\inst{\ref{aff31}}
\and J.~Skottfelt\orcid{0000-0003-1310-8283}\inst{\ref{aff104}}
\and L.~Stanco\orcid{0000-0002-9706-5104}\inst{\ref{aff96}}
\and P.~Tallada-Cresp\'{i}\orcid{0000-0002-1336-8328}\inst{\ref{aff40},\ref{aff41}}
\and A.~N.~Taylor\inst{\ref{aff45}}
\and H.~I.~Teplitz\orcid{0000-0002-7064-5424}\inst{\ref{aff105}}
\and I.~Tereno\orcid{0000-0002-4537-6218}\inst{\ref{aff106},\ref{aff107}}
\and S.~Toft\orcid{0000-0003-3631-7176}\inst{\ref{aff108},\ref{aff109}}
\and R.~Toledo-Moreo\orcid{0000-0002-2997-4859}\inst{\ref{aff110},\ref{aff111}}
\and F.~Torradeflot\orcid{0000-0003-1160-1517}\inst{\ref{aff41},\ref{aff40}}
\and I.~Tutusaus\orcid{0000-0002-3199-0399}\inst{\ref{aff112},\ref{aff113},\ref{aff99}}
\and J.~Valiviita\orcid{0000-0001-6225-3693}\inst{\ref{aff73},\ref{aff74}}
\and T.~Vassallo\orcid{0000-0001-6512-6358}\inst{\ref{aff15},\ref{aff56}}
\and G.~Verdoes~Kleijn\orcid{0000-0001-5803-2580}\inst{\ref{aff114}}
\and A.~Veropalumbo\orcid{0000-0003-2387-1194}\inst{\ref{aff20},\ref{aff33},\ref{aff32}}
\and Y.~Wang\orcid{0000-0002-4749-2984}\inst{\ref{aff97}}
\and J.~Weller\orcid{0000-0002-8282-2010}\inst{\ref{aff58},\ref{aff57}}
\and F.~M.~Zerbi\orcid{0000-0002-9996-973X}\inst{\ref{aff20}}
\and E.~Zucca\orcid{0000-0002-5845-8132}\inst{\ref{aff2}}
\and M.~Sereno\orcid{0000-0003-0302-0325}\inst{\ref{aff2},\ref{aff31}}}
										   
%%%% please do not edit the affiliation list -- contact ECEB Bureau for changes
\institute{CEA Saclay, DFR/IRFU, Service d'Astrophysique, Bat. 709, 91191 Gif-sur-Yvette, France\label{aff1}
\and
INAF-Osservatorio di Astrofisica e Scienza dello Spazio di Bologna, Via Piero Gobetti 93/3, 40129 Bologna, Italy\label{aff2}
\and
Universit\'e Paris-Saclay, Universit\'e Paris Cit\'e, CEA, CNRS, AIM, 91191, Gif-sur-Yvette, France\label{aff3}
\and
Universit\'e de Strasbourg, CNRS, Observatoire astronomique de Strasbourg, UMR 7550, 67000 Strasbourg, France\label{aff4}
\and
Center for Astronomy and Astrophysics and Department of Physics, Fudan University, Shanghai 200438, People's Republic of China\label{aff5}
\and
School of Physics \& Astronomy, University of Southampton, Highfield Campus, Southampton SO17 1BJ, UK\label{aff6}
\and
School of Physics and Astronomy, Anqing Normal University, Anqing 246133, China\label{aff7}
\and
Institute of Astronomy and Astrophysics, Anqing Normal University, Anqing 246133, China\label{aff8}
\and
Key Laboratory of Modern Astronomy and Astrophysics (Nanjing University), Ministry of Education, Nanjing 210093, China\label{aff9}
\and
Leiden Observatory, Leiden University, Einsteinweg 55, 2333 CC Leiden, The Netherlands\label{aff10}
\and
Universiteit Gent, Department of Physics and Astronomy, Proeftuinstraat 86 N3, 9000 Ghent, Belgium
\label{aff11}
\and
INAF-Osservatorio Astronomico di Padova, Via dell'Osservatorio 5, 35122 Padova, Italy\label{aff12}
\and
Instituto de Astrof\'{\i}sica de Canarias, E-38205 La Laguna, Tenerife, Spain\label{aff13}
\and
Universidad de La Laguna, Dpto. Astrof\'\i sica, E-38206 La Laguna, Tenerife, Spain\label{aff14}
\and
INAF-Osservatorio Astronomico di Trieste, Via G. B. Tiepolo 11, 34143 Trieste, Italy\label{aff15}
\and
Institute de Physique du Globe de Paris, 1 Rue Jussieu, 75005, Paris, France\label{aff16}
\and
SISSA, International School for Advanced Studies, Via Bonomea 265, 34136 Trieste TS, Italy\label{aff17}
\and
National Centre for Nuclear Research, ul. Pasteura 7, 02-093, Warsaw, Poland\label{aff18}
\and
Dipartimento di Fisica ``G. Occhialini", Universit\`a degli Studi di Milano Bicocca, Piazza della Scienza 3, 20126 Milano, Italy\label{aff19}
\and
INAF-Osservatorio Astronomico di Brera, Via Brera 28, 20122 Milano, Italy\label{aff20}
\and
Instituto de Fisica, Pontificia Universidad Catolica de Valparaiso, Valparaiso, Chile\label{aff21}
\and
INAF-Istituto di Astrofisica e Planetologia Spaziali, via del Fosso del Cavaliere, 100, 00100 Roma, Italy\label{aff22}
\and
Dipartimento di Fisica e Astronomia "Augusto Righi" - Alma Mater Studiorum Universit\`a di Bologna, via Piero Gobetti 93/2, 40129 Bologna, Italy\label{aff23}
\and
Univ. Lille, CNRS, Centrale Lille, UMR 9189 CRIStAL, 59000 Lille, France\label{aff24}
\and
Universit\'e Paris-Saclay, CNRS, Institut d'astrophysique spatiale, 91405, Orsay, France\label{aff25}
\and
IFPU, Institute for Fundamental Physics of the Universe, via Beirut 2, 34151 Trieste, Italy\label{aff26}
\and
Zentrum f\"ur Astronomie, Universit\"at Heidelberg, Philosophenweg 12, 69120 Heidelberg, Germany\label{aff27}
\and
ESAC/ESA, Camino Bajo del Castillo, s/n., Urb. Villafranca del Castillo, 28692 Villanueva de la Ca\~nada, Madrid, Spain\label{aff28}
\and
INFN, Sezione di Trieste, Via Valerio 2, 34127 Trieste TS, Italy\label{aff29}
\and
Dipartimento di Fisica e Astronomia, Universit\`a di Bologna, Via Gobetti 93/2, 40129 Bologna, Italy\label{aff30}
\and
INFN-Sezione di Bologna, Viale Berti Pichat 6/2, 40127 Bologna, Italy\label{aff31}
\and
Dipartimento di Fisica, Universit\`a di Genova, Via Dodecaneso 33, 16146, Genova, Italy\label{aff32}
\and
INFN-Sezione di Genova, Via Dodecaneso 33, 16146, Genova, Italy\label{aff33}
\and
Department of Physics "E. Pancini", University Federico II, Via Cinthia 6, 80126, Napoli, Italy\label{aff34}
\and
INAF-Osservatorio Astronomico di Capodimonte, Via Moiariello 16, 80131 Napoli, Italy\label{aff35}
\and
Dipartimento di Fisica, Universit\`a degli Studi di Torino, Via P. Giuria 1, 10125 Torino, Italy\label{aff36}
\and
INFN-Sezione di Torino, Via P. Giuria 1, 10125 Torino, Italy\label{aff37}
\and
INAF-Osservatorio Astrofisico di Torino, Via Osservatorio 20, 10025 Pino Torinese (TO), Italy\label{aff38}
\and
INAF-IASF Milano, Via Alfonso Corti 12, 20133 Milano, Italy\label{aff39}
\and
Centro de Investigaciones Energ\'eticas, Medioambientales y Tecnol\'ogicas (CIEMAT), Avenida Complutense 40, 28040 Madrid, Spain\label{aff40}
\and
Port d'Informaci\'{o} Cient\'{i}fica, Campus UAB, C. Albareda s/n, 08193 Bellaterra (Barcelona), Spain\label{aff41}
\and
INAF-Osservatorio Astronomico di Roma, Via Frascati 33, 00078 Monteporzio Catone, Italy\label{aff42}
\and
INFN -- Sezione di Napoli, Via Cinthia 6, 80126, Napoli, Italy\label{aff43}
\and
Dipartimento di Fisica e Astronomia "Augusto Righi" - Alma Mater Studiorum Universit\`a di Bologna, Viale Berti Pichat 6/2, 40127 Bologna, Italy\label{aff44}
\and
Institute for Astronomy, University of Edinburgh, Royal Observatory, Blackford Hill, Edinburgh EH9 3HJ, UK\label{aff45}
\and
European Space Agency/ESRIN, Largo Galileo Galilei 1, 00044 Frascati, Roma, Italy\label{aff46}
\and
Universit\'e Claude Bernard Lyon 1, CNRS/IN2P3, IP2I Lyon, UMR 5822, Villeurbanne, F-69100, France\label{aff47}
\and
Institut de Ci\`{e}ncies del Cosmos (ICCUB), Universitat de Barcelona (IEEC-UB), Mart\'{i} i Franqu\`{e}s 1, 08028 Barcelona, Spain\label{aff48}
\and
Instituci\'o Catalana de Recerca i Estudis Avan\c{c}ats (ICREA), Passeig de Llu\'{\i}s Companys 23, 08010 Barcelona, Spain\label{aff49}
\and
Institut de Ciencies de l'Espai (IEEC-CSIC), Campus UAB, Carrer de Can Magrans, s/n Cerdanyola del Vall\'es, 08193 Barcelona, Spain\label{aff50}
\and
UCB Lyon 1, CNRS/IN2P3, IUF, IP2I Lyon, 4 rue Enrico Fermi, 69622 Villeurbanne, France\label{aff51}
\and
Mullard Space Science Laboratory, University College London, Holmbury St Mary, Dorking, Surrey RH5 6NT, UK\label{aff52}
\and
Department of Astronomy, University of Geneva, ch. d'Ecogia 16, 1290 Versoix, Switzerland\label{aff53}
\and
Space Science Data Center, Italian Space Agency, via del Politecnico snc, 00133 Roma, Italy\label{aff54}
\and
School of Physics, HH Wills Physics Laboratory, University of Bristol, Tyndall Avenue, Bristol, BS8 1TL, UK\label{aff55}
\and
University Observatory, LMU Faculty of Physics, Scheinerstr.~1, 81679 Munich, Germany\label{aff56}
\and
Max Planck Institute for Extraterrestrial Physics, Giessenbachstr. 1, 85748 Garching, Germany\label{aff57}
\and
Universit\"ats-Sternwarte M\"unchen, Fakult\"at f\"ur Physik, Ludwig-Maximilians-Universit\"at M\"unchen, Scheinerstr.~1, 81679 M\"unchen, Germany\label{aff58}
\and
National Research Council, Herzberg Astronomy and Astrophysics Research Centre, 5071 W. Saanich Rd. Victoria, BC, V9E 2E7, Canada\label{aff59}
\and
Institute of Theoretical Astrophysics, University of Oslo, P.O. Box 1029 Blindern, 0315 Oslo, Norway\label{aff60}
\and
Jet Propulsion Laboratory, California Institute of Technology, 4800 Oak Grove Drive, Pasadena, CA, 91109, USA\label{aff61}
\and
Department of Physics, Lancaster University, Lancaster, LA1 4YB, UK\label{aff62}
\and
Felix Hormuth Engineering, Goethestr. 17, 69181 Leimen, Germany\label{aff63}
\and
Technical University of Denmark, Elektrovej 327, 2800 Kgs. Lyngby, Denmark\label{aff64}
\and
Cosmic Dawn Center (DAWN), Denmark\label{aff65}
\and
Universit\'e PSL, Observatoire de Paris, Sorbonne Universit\'e, CNRS, LERMA, 75014, Paris, France\label{aff66}
\and
Universit\'e Paris-Cit\'e, 5 Rue Thomas Mann, 75013, Paris, France\label{aff67}
\and
Max-Planck-Institut f\"ur Astronomie, K\"onigstuhl 17, 69117 Heidelberg, Germany\label{aff68}
\and
NASA Goddard Space Flight Center, Greenbelt, MD 20771, USA\label{aff69}
\and
Department of Physics and Astronomy, University College London, Gower Street, London WC1E 6BT, UK\label{aff70}
\and
Aix-Marseille Universit\'e, CNRS/IN2P3, CPPM, Marseille, France\label{aff71}
\and
Universit\'e de Gen\`eve, D\'epartement de Physique Th\'eorique and Centre for Astroparticle Physics, 24 quai Ernest-Ansermet, CH-1211 Gen\`eve 4, Switzerland\label{aff72}
\and
Department of Physics, P.O. Box 64, University of Helsinki, 00014 Helsinki, Finland\label{aff73}
\and
Helsinki Institute of Physics, Gustaf H{\"a}llstr{\"o}min katu 2, University of Helsinki, 00014 Helsinki, Finland\label{aff74}
\and
Laboratoire d'etude de l'Univers et des phenomenes eXtremes, Observatoire de Paris, Universit\'e PSL, Sorbonne Universit\'e, CNRS, 92190 Meudon, France\label{aff75}
\and
SKAO, Jodrell Bank, Lower Withington, Macclesfield SK11 9FT, UK\label{aff76}
\and
Centre de Calcul de l'IN2P3/CNRS, 21 avenue Pierre de Coubertin 69627 Villeurbanne Cedex, France\label{aff77}
\and
Universit\"at Bonn, Argelander-Institut f\"ur Astronomie, Auf dem H\"ugel 71, 53121 Bonn, Germany\label{aff78}
\and
INFN-Sezione di Roma, Piazzale Aldo Moro, 2 - c/o Dipartimento di Fisica, Edificio G. Marconi, 00185 Roma, Italy\label{aff79}
\and
Aix-Marseille Universit\'e, CNRS, CNES, LAM, Marseille, France\label{aff80}
\and
Department of Physics, Institute for Computational Cosmology, Durham University, South Road, Durham, DH1 3LE, UK\label{aff81}
\and
Universit\'e Paris Cit\'e, CNRS, Astroparticule et Cosmologie, 75013 Paris, France\label{aff82}
\and
CNRS-UCB International Research Laboratory, Centre Pierre Bin\'etruy, IRL2007, CPB-IN2P3, Berkeley, USA\label{aff83}
\and
University of Applied Sciences and Arts of Northwestern Switzerland, School of Engineering, 5210 Windisch, Switzerland\label{aff84}
\and
Institute of Physics, Laboratory of Astrophysics, Ecole Polytechnique F\'ed\'erale de Lausanne (EPFL), Observatoire de Sauverny, 1290 Versoix, Switzerland\label{aff85}
\and
Telespazio UK S.L. for European Space Agency (ESA), Camino bajo del Castillo, s/n, Urbanizacion Villafranca del Castillo, Villanueva de la Ca\~nada, 28692 Madrid, Spain\label{aff86}
\and
Institut de F\'{i}sica d'Altes Energies (IFAE), The Barcelona Institute of Science and Technology, Campus UAB, 08193 Bellaterra (Barcelona), Spain\label{aff87}
\and
European Space Agency/ESTEC, Keplerlaan 1, 2201 AZ Noordwijk, The Netherlands\label{aff88}
\and
DARK, Niels Bohr Institute, University of Copenhagen, Jagtvej 155, 2200 Copenhagen, Denmark\label{aff89}
\and
Waterloo Centre for Astrophysics, University of Waterloo, Waterloo, Ontario N2L 3G1, Canada\label{aff90}
\and
Department of Physics and Astronomy, University of Waterloo, Waterloo, Ontario N2L 3G1, Canada\label{aff91}
\and
Perimeter Institute for Theoretical Physics, Waterloo, Ontario N2L 2Y5, Canada\label{aff92}
\and
Centre National d'Etudes Spatiales -- Centre spatial de Toulouse, 18 avenue Edouard Belin, 31401 Toulouse Cedex 9, France\label{aff93}
\and
Institute of Space Science, Str. Atomistilor, nr. 409 M\u{a}gurele, Ilfov, 077125, Romania\label{aff94}
\and
Dipartimento di Fisica e Astronomia "G. Galilei", Universit\`a di Padova, Via Marzolo 8, 35131 Padova, Italy\label{aff95}
\and
INFN-Padova, Via Marzolo 8, 35131 Padova, Italy\label{aff96}
\and
Caltech/IPAC, 1200 E. California Blvd., Pasadena, CA 91125, USA\label{aff97}
\and
Instituto de F\'isica Te\'orica UAM-CSIC, Campus de Cantoblanco, 28049 Madrid, Spain\label{aff98}
\and
Institut de Recherche en Astrophysique et Plan\'etologie (IRAP), Universit\'e de Toulouse, CNRS, UPS, CNES, 14 Av. Edouard Belin, 31400 Toulouse, France\label{aff99}
\and
Universit\'e St Joseph; Faculty of Sciences, Beirut, Lebanon\label{aff100}
\and
Departamento de F\'isica, FCFM, Universidad de Chile, Blanco Encalada 2008, Santiago, Chile\label{aff101}
\and
Universit\"at Innsbruck, Institut f\"ur Astro- und Teilchenphysik, Technikerstr. 25/8, 6020 Innsbruck, Austria\label{aff102}
\and
Department of Physics and Helsinki Institute of Physics, Gustaf H\"allstr\"omin katu 2, University of Helsinki, 00014 Helsinki, Finland\label{aff103}
\and
Centre for Electronic Imaging, Open University, Walton Hall, Milton Keynes, MK7~6AA, UK\label{aff104}
\and
Infrared Processing and Analysis Center, California Institute of Technology, Pasadena, CA 91125, USA\label{aff105}
\and
Departamento de F\'isica, Faculdade de Ci\^encias, Universidade de Lisboa, Edif\'icio C8, Campo Grande, PT1749-016 Lisboa, Portugal\label{aff106}
\and
Instituto de Astrof\'isica e Ci\^encias do Espa\c{c}o, Faculdade de Ci\^encias, Universidade de Lisboa, Tapada da Ajuda, 1349-018 Lisboa, Portugal\label{aff107}
\and
Cosmic Dawn Center (DAWN)\label{aff108}
\and
Niels Bohr Institute, University of Copenhagen, Jagtvej 128, 2200 Copenhagen, Denmark\label{aff109}
\and
Universidad Polit\'ecnica de Cartagena, Departamento de Electr\'onica y Tecnolog\'ia de Computadoras,  Plaza del Hospital 1, 30202 Cartagena, Spain\label{aff110}
\and
European University of Technology EUt+, European Union\label{aff111}
\and
Institute of Space Sciences (ICE, CSIC), Campus UAB, Carrer de Can Magrans, s/n, 08193 Barcelona, Spain\label{aff112}
\and
Institut d'Estudis Espacials de Catalunya (IEEC),  Edifici RDIT, Campus UPC, 08860 Castelldefels, Barcelona, Spain\label{aff113}
\and
Kapteyn Astronomical Institute, University of Groningen, PO Box 800, 9700 AV Groningen, The Netherlands\label{aff114}}

 \date{\today}

% 
% Put your abstract here
%
 \abstract{We present the discovery of two disky titans {in the first data release of the \textit{Euclid} satellite}. These sources are massive ($M_\ast>10^{11} \, M_\odot $) star-forming (${\rm SFR}\sim 20 \, M_\odot\,{\rm yr}^{-1}$) disks located in strong over-dense environments at intermediate redshifts ($z\sim0.75$). They represent a small fraction of the massive galaxies in over-dense regions (just four candidates in over $20\,\mathrm{deg}^2$ {analysed in this study}). Their existence is puzzling, given the abundance of passive, bulge-dominated sources commonly found at the centre of groups and clusters at low redshifts. First, our analysis shows that these objects are located in massive groups ($M_{\rm h}\sim10^{13.8} \, M_\odot$), where rapid accretion of cold gas should be prevented by the formation of a static hot halo. Despite this, a millimetre follow-up with NOEMA shows significant cold gas reservoirs ($M_{\rm H_2}\sim10^{10.3} \, M_\odot$) within these sources. Second, our morphological analysis indicates a massive, passive bulge in these galaxies. This bulge is expected to stabilise the disk against fragmentation, thereby suppressing further star formation. However, these sources lie on the Schmidt--Kennicutt relation or even slightly above. Building on these observations, we propose a scenario where these disky titans are the product of a merger-induced rejuvenation episode, in which the most massive galaxy in a group accretes cold gas from another member and briefly restarts star formation. Such a scenario is supported by a comparison with the TNG300 simulation and easily explains the persistence of star formation activity in massive galaxies in over-dense environments as temporary stages of more complex evolution. More generally, our study showcases the ability of \textit{Euclid} to find rare objects, thanks to the unprecedented statistic offered by its surveys and the scientific potential residing in the synergy between \textit{Euclid} and other facilities observing at longer wavelengths.}

%
% Provide up to five key words:
%
\keywords{galaxies:evolution --- galaxies:halos --- galaxies:ISM --- galaxies:spiral}
%
% Add short versions of title and author list for page headings
%
   \titlerunning{Disky titans with \Euclid}
   \authorrunning{F. Gentile et al.}
   
   \maketitle
%
%-------------------------------------------------------------------
%

\section{Introduction}
\label{sec:intro}

One of the main observational results in extragalactic astronomy is the strong bimodality between star-forming and quiescent galaxies \citep[e.g.][]{Kauffmann_03,Brichmann_04}. The first population is composed of objects with blue rest-frame colours, high star formation activity, and younger stellar ages, whereas the second includes galaxies with redder colours, lower star formation rates (SFRs), and older stellar populations. Quite remarkably, this distinction extends to other physical properties such as morphology: red objects are bulge-dominated, while blue sources exhibit more dominant disky structures (see e.g. \citealt{Wuyts_11,Q1-SP040}), with notable exceptions available in the literature (see e.g. \citealt{Masters_10,McIntosh_14}). The existence of this bimodality is commonly interpreted as evidence that two main processes -- galaxy quenching and morphological transformation -- must occur during the evolution of galaxies. Quenching strongly suppresses star formation activity, while morphological transformation involves the transition from a disk-dominated to a bulge-dominated morphology. Moreover, the small number of galaxies found in between these two main populations is commonly interpreted as evidence that both phenomena occur on timescales much shorter than the {total time galaxies spend as either star-forming or passive.} Despite decades of study by the community, a clear picture of these two phenomena, their interplay, and the physical mechanisms responsible for them remains missing.

From an observational perspective, both phenomena are tightly related to the galaxy stellar mass (with the fraction of passive and spheroidal objects in a given environment consistently higher at higher masses; e.g. \citealt{Pozzetti_10,Shuntov_25}), and to their local environment (the same fractions increase in denser environments at fixed stellar mass; e.g. \citealt{Dressler_80,Bolzonella_10,Peng_10,Q1-SP017,Q1-SP069}). For this reason, we commonly introduce two large families of processes: mass quenching and environmental quenching (see e.g. \citealt{Peng_10}). The mass quenching group includes all internal processes commonly associated with higher stellar masses, such as feedback from active galactic nuclei (AGNs; e.g. \citealt{Bower_06,Croton_06}) or morphological quenching \citep{Martig_09,Martig_13}. In the latter, a massive and passive bulge (more common in massive galaxies) stabilises the stellar disk, preventing star formation. The other group includes external mechanisms that are thought to be more efficient in dense environments either by preventing the accretion of cold gas onto galaxies (e.g. through halo shock heating or starvation, mostly affecting high-mass sources; \citealt{Dekel_06, Van_Den_Bosch_2008}) or by removing it (e.g. through tidal stripping or ram-pressure stripping, which are more effective in low-mass objects; \citealt{Gunn_72,Feldmann_10,Boselli_22}). Both families of processes are thus expected to show their maximum efficiency in the massive galaxies located at the centre of over-dense regions (typically referred to as brightest group or cluster galaxies  -- BGGs or BCGs -- depending on the size of the structures hosting them). Indeed, these sources are found -- in the vast majority of cases -- to be passive and spheroidal galaxies, at least up to $z\sim1$ (e.g. \citealt{Lopez_Cruz_04,Whiley_08,Fraser_McKelvie_14}), even though rare exceptions have been discovered and well studied in the past decade (see e.g. \citealt{Fogarty_15,Fogarty_19,Castignani_20,Castignani_22,Castignani_23}).

The overall scenario characterised by the interplay between the mass and environmental quenching outlined above has been confirmed in the last decade by several observational studies and simulations (e.g. \citealt{Peng_10,Bahe_17,Henriques_17,Weller_25,Ghaffari_26}). In the last year, in particular, the launch of the \Euclid space telescope gave us access to an unprecedented statistic of galaxies, including reliable estimates of their physical properties and morphologies, coupled with a well-characterised environment. These new possibilities have already allowed us to confirm the overall correlation between properties, morphologies, mass, and environment \citep{Q1-SP017}, as well as separate the study of quenching and morphological transformation and analyse their interplay \citep{Q1-SP069}.

In this paper, we focus on another main asset of \Euclid: its ability to find rare objects. A large statistical sample is thus crucial to find outliers of well-assessed relations, giving us the opportunity to investigate in detail cases in which the processes thought to drive the observed trends seem less efficient or entirely absent. Specifically, in this paper we present the discovery of two 'disky titans'. These rare sources (just four in over 20 deg$^2$ covered by the \Euclid telescope in the first data released by the collaboration) are defined as massive ($M_\ast>10^{11} \, M_\odot$), disky, star-forming (${\rm SFR}\gtrsim 20 \, M_\odot\, {\rm yr}^{-1}$) galaxies, despite their location in strong over-dense regions. 

This paper focuses on the characterisation of a first (pilot) sample of two sources, for which millimetre (mm) data obtained from the Northern Extended Millimetre Array (NOEMA) are available. The main goal of this study is to constrain their physical properties, characterise their environment, model their morphology, and offer a likely scenario explaining their observed properties.

This paper is structured as follows. In Sect.\,\ref{sec:data} we present the multi-wavelength data on which our analysis is based. Section\,\ref{sec:prop_gal} focuses on the characterisation of the two galaxies in terms of gas content, physical properties, and morphology. In Sect.\,\ref{sec:prop_groups}, we characterise the two groups in which the two galaxies are located and estimate their halo mass. In Sect.\,\ref{sec:discussion} we discuss our findings, propose a likely scenario to explain our observations, and verify its plausibility by looking for analogue sources in hydrodynamical simulations. Finally, we draw our conclusions in Sect.\,\ref{sec:summary} and forecast the possible future perspective of this work.

Throughout this study, we assume a standard $\Lambda$CDM cosmology with the parameters reported in \citet{Planck_18}. We also assume the initial mass function by \citet{Chabrier_03}. Finally, all the magnitudes quoted throughout the text are reported in the AB photometric system \citep{Oke_83}.

\begin{figure*}
    \centering
    \includegraphics[width=\linewidth]{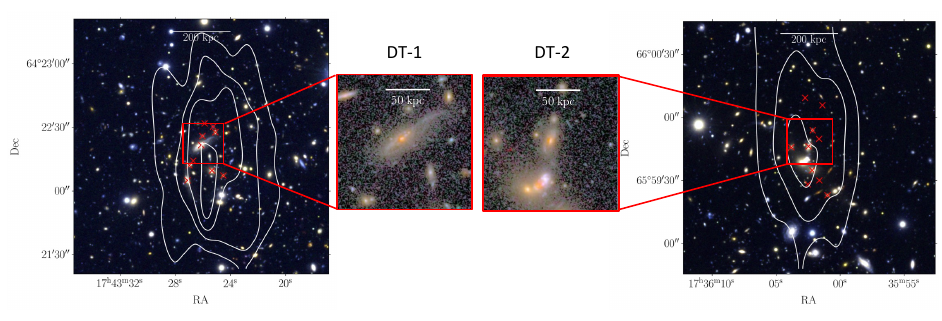}
    \caption[]{Colour-composite images of the two targets analysed in this paper and the surrounding environments. Each large image covers one square arc minute, with a side roughly corresponding to $441$ proper $\mathrm{kpc}$ at $z\sim0.7$. These images were obtained through the \citet{Lupton_01} algorithm by combining the \IE, \YE, and \HE images. The white curves represent the galaxy density field in units of $\sigma$ above the median, starting from $3\sigma$. The candidate members (i.e. with a redshift consistent with the over-density and a density contrast exceeding $3\sigma$) are marked with red crosses. Inner panels: Zoom-in on the two disky titans, with sides of $20\arcsec \times 20\arcsec$. These images are generated using \texttt{Jafar}\footnotemark\ to highlight low-surface-brightness features by combining the four \Euclid bands.}
    \label{fig:groups}
\end{figure*}

\section{Data and target selection}
\label{sec:data}

\subsection{Euclid data}
\label{sec:euclid}
The space-based imaging and photometry analysed in this paper are from the \citet[DR1;][]{DR1cite} of the \Euclid telescope. These data include a first small fraction (about $1900\,\mathrm{deg}^2$) of the Euclid Wide Survey \citep[EWS;][]{Scaramella-EP1} and deeper observations of the Euclid Deep Fields (EDFs; about $53\,\mathrm{deg}^2$) already included in the first Quick Data Release (\citealt{Q1cite,Q1-TP001}). Unlike Q1 data, DR1 includes several visits covering two of the EDFs (Euclid Deep Field North and South, EDF-N and EDF-S, respectively) to reach higher depths.

\Euclid provides space-based imaging in four filters: one large optical band (\IE, \citealt{EuclidSkyVIS}) and three near-infrared ones (\YE, \JE, and \HE; \citealt{EuclidSkyNISP}). The two targets presented in this paper are located in EDF-N (about 22.9 deg$^2$; Euclid Collaboration: Aussel et al., \textit{in prep.}). Hence, they are covered by ten passes of the telescope released in DR1, ensuring a $5\sigma$ depth in $2\arcsec$ apertures of 26.5, 24.9, 25.1, and 25.0 in \IE, \YE, \JE, and \HE, respectively (about 1 magnitude deeper than Q1 data; \citealt{Q1-TP004}). 

These space-based data are complemented by ground-based observations, collected as part of the Ultraviolet Near Infrared Optical Northern Survey (UNIONS; \citealt{Gwyn_25}). This survey grants ancillary data in the five $ugriz$ bands by combining different surveys observing the northern hemisphere, with depths spanning from $23.1\mbox{--}24.9\,\mathrm{mag}$. Further details on the specific surveys used for ground-based ancillary photometry can be found in \citet{EuclidSkyOverview} and in \citet{Gwyn_25}.

Photometry was extracted in all bands through the MERge (MER) processing function (responsible for assembling the photometric catalogues from both the \textit{Euclid} observations and external data; see Euclid Collaboration: Kümmel et al., in prep.) and passed to the photo-\textit{z} (PHZ) pipeline (Euclid Collaboration: Hartley et al., in prep.; Euclid Collaboration: Enia et al., in prep.). Together with this latter pipeline, photometric redshifts and physical properties were then derived using {the machine-learning algorithm, Nearest-Neighbour Photometric Redshift (\texttt{nnpz}; \citealt{Desprez-EP10, EP-Enia})}.

These measurements were used to compute a mass completeness limit for our data. Following the standard procedure by \citet{Pozzetti_10} applied to passive galaxies (colour-selected in the rest-frame $\mathrm{NUV}-r$, $r-J$ plane; \citealt{Ilbert_10}) brighter than $\HE<25$, we find that the EDF-N is 90\% mass-complete for masses higher than $M_{\rm lim}=10^{9} \, M_\odot$ at $z\sim0.8$. Since one of our targets is located in the fraction of EDF-N covered by \textit{Spitzer} observations as part of the Cosmic Dawn Survey \citep{Moneti-EP17,EP-McPartland}, we supplemented the \Euclid and external photometry with data from the Infrared Array Camera (IRAC), as extracted by \citet{Q1-SP011}.

\subsection{Target selection}
\label{sec:sample_selection}

The targets analysed in this paper were selected starting from the density maps built by \citet{Q1-SP069}. In this study, the authors employed the first Q1 data from \Euclid to model the density fields of the three EDFs in the redshift range $0.25<z<1$. Starting from their catalogue of EDF-N (accessible with ground-based facilities observing the northern sky and covered by deeper Euclid DR1 data), we first selected galaxies in strong over-densities, with a density contrast parameter $\log_{10}(1+\delta)$ exceeding $5\sigma$ above the median at their redshift. To reduce the contamination by satellite galaxies, we further selected galaxies with $M_\ast>10^{11} \, M_\odot$. Following these criteria, we obtained a sample of 42 likely BGGs or BCGs. {From this parent sample, we selected star-forming galaxies based on their $\mathrm{NUV}-r$ and $r-J$ colours \citep{Ilbert_10} and on their distance from the main sequence (MS) of star-forming galaxies (\citealt{Q1-SP031}), which are lower than the intrinsic scatter of the relation ($\sigma\sim0.3 \, {\rm dex}$)}. This additional step resulted in a sample of four star-forming BCG or BGG candidates in EDF-N. The two final candidate disky titans (hereafter dubbed DT-1 and DT-2) were selected based on the clear presence of a stellar disk {as confirmed by visual inspection} and of {their narrow redshift probability function} ($z_{\rm phot,1}=0.69\pm0.05$ and $z_{\rm phot,2}=0.78\pm0.01$, respectively). Colour-composite images of the two targets discussed in this paper and of the surrounding environments are shown in Fig.\,\ref{fig:groups} {while images of the two additional objects are reported in Appendix\,\ref{app:xrays}}. 

\footnotetext{\url{https://jafar.astro.unistra.fr/}}

\subsection{NOEMA data}
\label{sec:noema}

The mm data for the two objects analysed in this paper are from the NOEMA programme S25BT (P.I.: F. Gentile). These data consist of interferometric observations at $\lambda\sim2\,{\rm mm}$, with the main goal of observing the CO(2-1) line at $z\sim0.7$. The large bandwidth offered by the NOEMA interferometer is crucial to identify the targeted line when only photometric redshifts are available. The data were collected with the interferometer in its most compact configuration (D), ensuring a clean beam of $\ang{;;3.3} \times \ang{;;2.8}$ at the declination of our targets. Each target was observed for a total of eight hours, allowing us to achieve a noise level of $130 \, \mu{\rm Jy \, beam^{-1}}$ when integrated over $500 \, \kms$, slightly variable with frequency.

The data were reduced with the \textsc{gildas} code suite (version sep25a; \citealt{GILDAS}) by employing the standard NOEMA pipeline. The reduced data are composed of two data cubes, centred on the two targets, covering two spectral windows of nearly $8\,\mathrm{GHz}$ each in the range $134\mbox{--}142\mbox{/}149\mbox{--}157$ and $127\mbox{--}135\mbox{/}142\mbox{--}150\,\mathrm{GHz}$ for the two objects, respectively.

\subsection{X-ray data for DT-2}
\label{sec:xrays}

DT-2 was serendipitously observed by the X-ray Multi-Mirror Mission (XMM-Newton) telescope, as part of the programme 0202100301 (P.I.: A. Wolter). The observations targeted the AGN NEP 2131 \citep{Wolter_05}, located around $7\,\mathrm{arcmin}$ from DT-2, with an exposure time of $\num{37409}\,{\rm s}$. It is interesting to note how the structure hosting DT-2 was already listed as a possible cluster of galaxies by \citet{Mehrtens_12} due to its extended X-ray emission in the soft band. Unfortunately, the lack of deep optical observations before \Euclid made it impossible to perform a complete analysis of this object. We downloaded the raw X-ray data from the XMM archive\footnote{\url{https://nxsa.esac.esa.int/nxsa-web/\#search}} and performed the reduction. {We employed the Science Analysis System (SAS v21.1\footnote{\url{https://www.cosmos.esa.int/web/xmm-newton/download-and-install-sas}}) -- a suite of libraries expecially designed to reduce and analyse data collected by XMM -- for data reduction, image creation, and spectral extraction, following standard procedures. In summary, we first used the \texttt{epproc} and \texttt{emproc} tools to concatenate and calibrate the event lists. Then, we employed \texttt{evselect} to create light-curves in the $10\mbox{--}12\,\mathrm{keV}$ and $>10\,\mathrm{keV}$ energy bands for the PN and MOS instruments, respectively. We filtered out background flaring periods with count rates greater than $0.4$, $0.15$, and $0.25\,{\rm cts}\,{\rm s}^{-1}$ for the PN, MOS1, and MOS2 cameras, respectively, resulting in clear exposure times of 18.0, 21.7, $21.6\,{\rm ks}$. Finally, we used the \texttt{evselect}, \texttt{eexpmap}, \texttt{backscale}, \texttt{rmfgen}, and \texttt{arfgen} tools to create images and exposure maps and to extract spectra, response matrices, and ancillary files.} These data are used in Sect.\,\ref{sec:halo_mass} to constrain the halo mass of the structure hosting DT-2. {No observations covering DT-1 were found in the archive.}

\section{Characterisation of the galaxies}
\label{sec:prop_gal}

\begin{table}[]
    \centering
    \caption{Main properties estimated for the two disky titans presented in this paper. }
    \label{tab:prop_gal}
    \begin{tabular}{lcc}
     & DT-1 & DT-2 \\
    \noalign{\vskip 1pt}
    \hline
    \noalign{\vskip 1pt}
    RA & \ra{17;43;26.17} & \ra{17;36;02.45} \\
    Dec & \ang{+64;22;21.50} & \ang{+65;59;46.05} \\ 
    $z_{\rm spec}$  & $0.6990 \pm 0.0002$ & $0.7672 \pm 0.0001$ \\ 
    \noalign{\vskip 1pt}
    \hline
    \noalign{\vskip 1pt}
    $\log_{10}(M_\ast/M_\odot)$  & $11.30\pm0.05$ & $11.25\pm0.08$ \\ 
    $\log_{10}({\rm SFR_{SED}} /M_\odot \ {\rm yr}^{-1})$ & $1.4\pm0.2$ & $1.4\pm0.2$ \\ 
    $\log_{10}({\rm SFR_{IR}}/M_\odot \ {\rm yr}^{-1}) $  & $1.3\pm0.1$ & $1.6\pm0.1$ \\ 
    $A_{\rm v}$ [mag]  & $2.1\pm0.6$ & $1.9\pm0.8$ \\ 
    \noalign{\vskip 1pt}
    \hline
    \noalign{\vskip 1pt}
    $r_{\rm phys}$ ($\IE$) [kpc] & $21.7\pm0.3$ & $13.5\pm0.1$ \\
    $r_{\rm phys}$ (${\rm H}_2$) [kpc] & $35\pm5$ & $25\pm2$ \\
    \noalign{\vskip 1pt}
    \hline
    \noalign{\vskip 1pt}
    $I_{\rm CO (2-1)}$ [Jy\,\kms] & $0.8\pm0.1$ & $0.83\pm0.09$ \\
    Width [\kms] & 442 & 689\\
    $\log_{10}(M_{\rm H_2}/M_{\odot})$ & $10.3 \pm 0.1$ & $10.4 \pm 0.1$\\
    $\mu_{\rm H_2}$ & $0.10\pm0.02$ & $0.14\pm0.04$ \\
    $\tau_{\rm dep}$ [Myr] & $794\pm375$ & $1000\pm690$ \\
    \noalign{\vskip 1pt}
    \hline
    \noalign{\vskip 1pt}
    $S_{\rm 2 mm}$ [$\mu$Jy] & $42\pm15$  & $86\pm24$ \\
    $\log_{10}(M_{\rm Dust}/M_\odot)$ & $8.0\pm0.3$ & $8.3\pm0.2$ \\
    $\log_{10}(M_{\rm Dust}/M_\ast)$ & $-3.3\pm0.4$ & $-3.0\pm0.3$ \\
    \noalign{\vskip 1pt}
    \hline
    \end{tabular}
    \tablefoot{The spectroscopic redshift of the galaxies, as well as the information about the gas content (Sect.\,\ref{sec:spec_gas}), is derived from the CO(2-1) transition observed with NOEMA. The constraints on the dust content are from the continuum emission, while the other physical properties were measured through SED fitting with \texttt{CIGALE} (Sect.\,\ref{sec:sed_fitting}).}

\end{table}

\subsection{Spec-z and gas mass}
\label{sec:spec_gas}

\begin{figure}
    \centering
    \includegraphics[width=\linewidth]{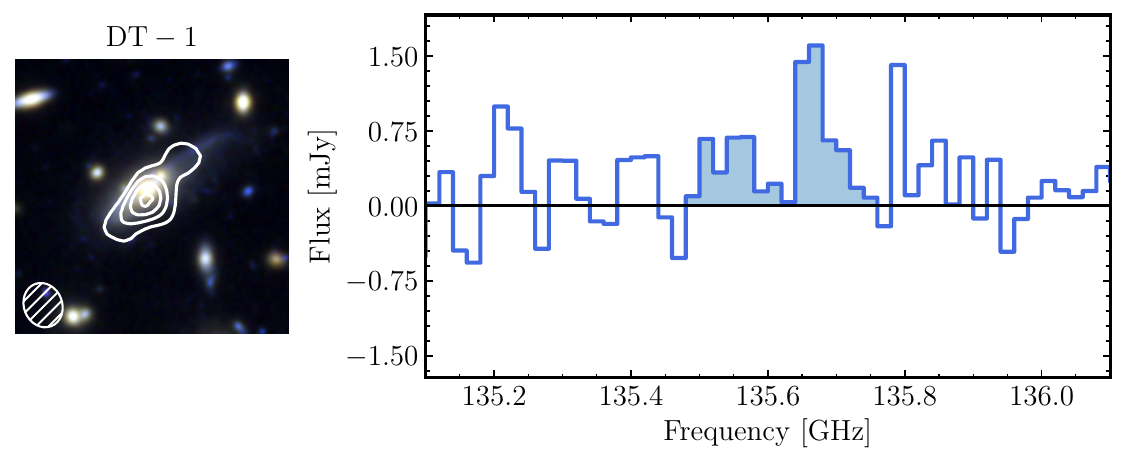} \\
    \includegraphics[width=\linewidth]{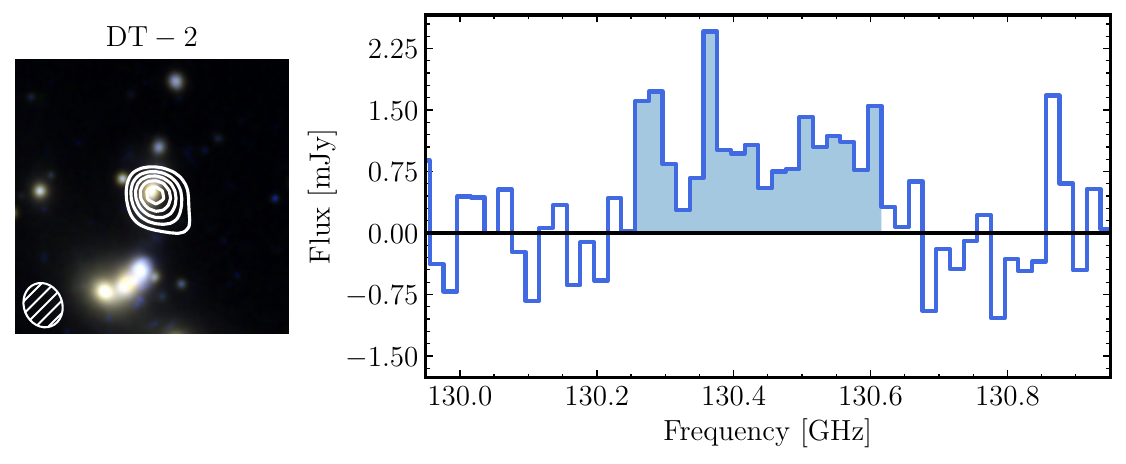}
    \caption{Two CO(2-1) lines observed with NOEMA in the two targets presented in this paper. \emph{Right panels}: Millimetre spectrum of each galaxy, with the channels belonging to the line (i.e. maximising its S/N; see Sect.\,\ref{sec:noema}) coloured in blue. \emph{Left panels}: Moment-zero of the line superimposed on the colour-composite image of the target. The contours increase by $1\sigma$, starting from $2\sigma$ and $5\sigma$ for the two galaxies, respectively. The synthesised beam is shown in the bottom left corner for reference.}
    \label{fig:noema}
\end{figure}

\begin{figure*}
    \centering
    \includegraphics[width=\linewidth]{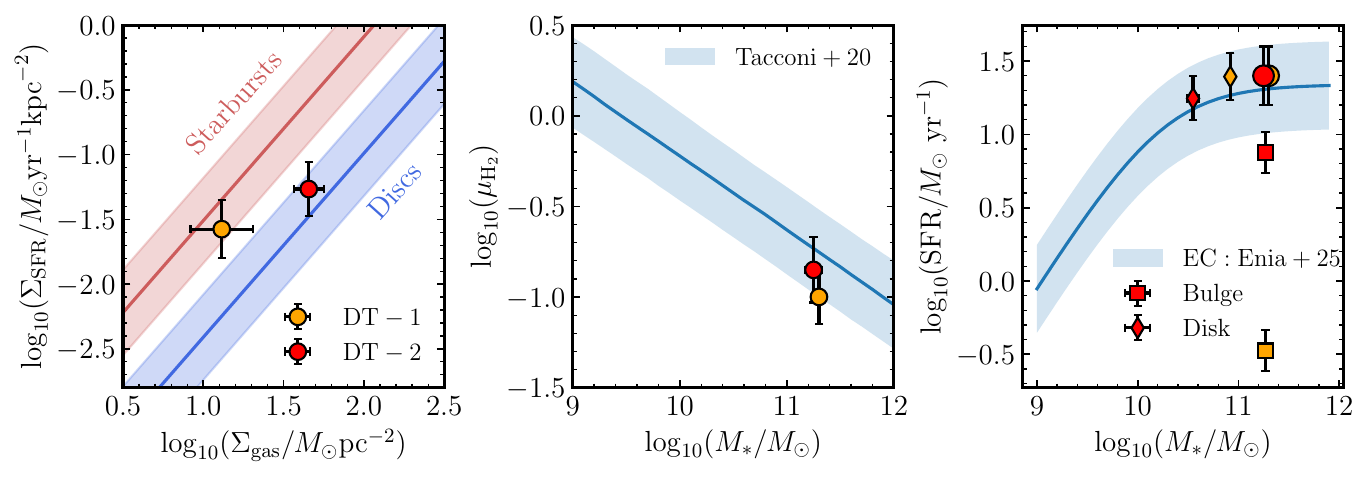}
    \caption{Comparison of two disky titans using the Schmidt--Kennicutt relation (\citealt{Schmidt_59,Kennicutt_98}; \textit{left panel}), the MS of star-forming galaxies (as parametrised by \citealt{Q1-SP031}; \textit{right panel}), and the molecular gas fraction expected for MS galaxies at $z\sim0.75$ (\citealt{Tacconi_20}; \textit{central panel}). In the \textit{left panel}, we report the sequence of disks and star-bursting galaxies following \citet{Daddi_10a}. In the \textit{right panel}, we report the masses and SFRs integrated over the entire galaxies, as well as for their bulges and disks separately (squares and diamonds, respectively), {derived following the procedure described in Sect.\,\ref{sec:morph}--\ref{sec:sed_fitting}.}}
    \label{fig:comparison}
\end{figure*}

An analysis of the two data cubes produced by NOEMA was performed using the Common Astronomy Software Applications (\texttt{CASA 6.7.2}; \citealt{CASA}) package. We followed the procedure described in \citet{Jin_19} and \citet{Gentile_24} to blindly search for emission line in our data. First, we extracted a spectrum at the expected position of our targets based on the high-resolution images provided by \Euclid. The extraction was performed in the visibility space with an elliptical Gaussian model with shape parameters obtained by fitting the model to the zero-moment of the CO emission. The extracted spectrum was then convolved with a series of boxcar filters with different widths, each time looking at the signal-to-noise ratio (S/N) of the brightest peak of emission. This procedure allows us to detect a bright line in each spectrum, with observed frequencies $\nu_1=(135.69\pm0.02)\,{\rm GHz}$ and $\nu_2=(130.45\pm0.01) \,{\rm GHz}$ and a S/N of 5.2 and 9.1 for DT-1 and DT-2, respectively. Following the procedure described in \citet{Jin_19}, we estimated a spurious probability of $10^{-3}$ and $10^{-6}$ for the two lines, sufficient to ensure the reliability of our detections. After masking the two detected lines, we stacked the remaining channels to detect a continuum signal, yielding two marginal detections with S/N values of 2.8 and 3.6, respectively. The observed lines, together with their $0$-moments are shown in Fig.\,\ref{fig:noema}. Even though the resolution does not allow for a complete dynamical modelling of the two lines, both exhibit a double-peaked profile -- a typical indication of rotation in the gas phase of our galaxies (as expected, given the disky shape of the stellar component).

The integrated flux ($I_{\rm CO(2-1)}$) {was computed by integrating the signal over the range of channels maximising the S/N. This latter quantity represents the width of the detected lines. These quantities are reported, together with the detected continua ($S_{\rm 2mm}$), in Table \ref{tab:prop_gal}}. Given the photometric redshifts of the two sources (Sect.\,\ref{sec:sample_selection}), we modelled the two lines as CO(2-1) transition ($\nu_0=230.538$ GHz), translating into spectroscopic redshifts of $z_1=0.6990 \pm 0.0002$ and $z_2=0.7672\pm 0.0001$, respectively. We emphasize that alternative models for the detected lines, although possible, would yield redshifts in strong tension with the observed photometry of the two targets.

Moreover, since the low rotational transitions of the CO molecule are well-known tracers of molecular gas (see e.g. the review by \citealt{Bolatto_13} and references therein), we employed the detected lines to estimate the gas mass of our targets. First, we rescaled the measured CO(2-1) fluxes of the two targets to the values expected for CO(1-0), assuming a standard ratio $R_{21}=0.64 \pm 0.09$ commonly adopted for star-forming galaxies (\citealp{denBrok_21}; see also another consistent estimate in \citealt{Daddi_15}). Then, we converted the CO(1-0) luminosity into a gas mass through the standard relation (e.g. \citealt{Bolatto_13})
\begin{equation}
    M_{\rm H_2}=3.25\times10^7\,\alpha_{\rm CO}\,I_{\rm CO(1-0)}\,\nu_{\rm obs}^{-2}\,D_{\rm L}^2\,(1+z)^{-3} \ ,
\end{equation}
where $D_{\rm L}$ is the luminosity distance.
Since, according to the selection scheme, the two targets are located on the MS (see Sect.\,\ref{sec:sample_selection}), we adopted a standard value of $\alpha_{\rm CO}=(3.6 \pm 0.8)\,M_\odot\,{\rm pc}^{-2}\,({\rm K\,\kms})^{-1}$, as commonly done for these types of sources \citep{Daddi_10a,Daddi_10b}. The estimated molecular gas masses, together with the gas fractions, $\mu_{\rm H_2}=M_{\rm H_2}/M_\ast$, and the depletion times, $\tau_{\rm dep}=M_{\rm H_2}/{\rm SFR}$, are reported in Table \ref{tab:prop_gal}. {We emphasize that choosing a lower value of $\alpha_{\rm CO}$ produces lower values of the gas mass and depletion time. For instance, a starburst value of $\alpha_{\rm CO}\sim0.8\,M_\odot\,{\rm pc}^{-2}\,({\rm K\,\kms})^{-1}$ (e.g. \citealt{Bolatto_13}) produces a gas mass that is 4.5 times lower. Such a chosen value, however, is not consistent with the observed SFR (see Sect.\,\ref{sec:sed_fitting}).

In the central panel of Fig.\,\ref{fig:comparison}, we compare the obtained molecular gas fraction with those expected for MS galaxies \citep{Tacconi_20} with the same mass at a reference redshift of $z=0.75$. We show that both estimates are consistent -- within the uncertainties -- with the expectations. Similarly, in the left panel of Fig.\,\ref{fig:comparison}, we compare the surface density of the molecular gas and SFR with the relation by \citet{Schmidt_59} and \citet{Kennicutt_98}, which typically holds for star-forming galaxies. To do this, we divided the two quantities by a reference surface density: that of the CO emission and that of the \IE emission for the molecular gas and SFR. We show that DT-2 lies -- within the uncertainties -- on the relation, while DT-1 is located above, indicating a lower depletion time and, consequently, a higher star formation efficiency (${\rm SFE}\sim\tau_{\rm dep}^{-1}$), consistent with the expectations for star-burst galaxies \citep{Daddi_10a}. These results are discussed in Sect.\,\ref{sec:discussion} to explain the observed properties of our two disky titans.

\subsection{Morphological analysis}
\label{sec:morph}

We performed a morphological analysis of our two targets by fitting two Sérsic profiles to their images at different wavelengths. {The fitting was performed through an Markov Chain Monte Carlo (MCMC) implemented with the \texttt{emcee} library \citep{Emcee_13}. We convolved the double Sérsic models with a set of Gaussian point-spread functions (PSFs) -- one for each band -- with the full width half maximum (FWHM) values equal to those reported in \citet{Q1-SP031}.} Each combined model has, in principle, $12$ free parameters, but we imposed equality between the centres of the two profiles and fixed the Sérsic index of the disk component to one, reducing the number of parameters to be fitted to nine. The combined model was then fitted to the four high-resolution images (\IE, \YE, \JE, and \HE) separately. We find that the bulge-to-total luminosity ratio strongly depends on the wavelength, increasing from 0.2 and 0.3 in the \IE band to 0.5 and 0.9 in \HE for the two targets, respectively. In addition, the two bulge components of the two galaxies in the near infrared (NIR) bands have a high Sérsic index ($n_{\rm bulge}\sim6$), consistent with a velocity-dispersion-dominated structure \citep{Kormendy_04,Fisher_08}. Figure\,\ref{fig:morph} shows the results of our modelling procedure. From the same figure (see also Fig.\,\ref{fig:groups}), it is also evident that both targets exhibit irregularities in their brightness profile: a significant tidal tail is visible in both sources, with a more pronounced feature present in DT-1. Similarly, a clumpy structure is visible in the residuals of DT-1, with three components detected in \IE and a single component visible in \HE. The presence of these features are used in Sect.\,\ref{sec:discussion} to constrain our evolutionary scenario for these sources. 

As a final step, to reconstruct the total spectral energy distribution (SED) of the two morphological components for our objects, we extended the complete analysis to the ground-based bands. Given the lower spatial resolution of these data, we fixed all the parameters of the double Sérsic profile to the values obtained on the \IE band (i.e. that with the highest wavelength overlap with those bands) with the sole exception of the normalisation. The obtained fluxes are employed in Sect.\,\ref{sec:sed_fitting} to estimate the physical properties of the two components of our galaxies.

\begin{figure*}
    \centering
    \includegraphics[width=\linewidth]{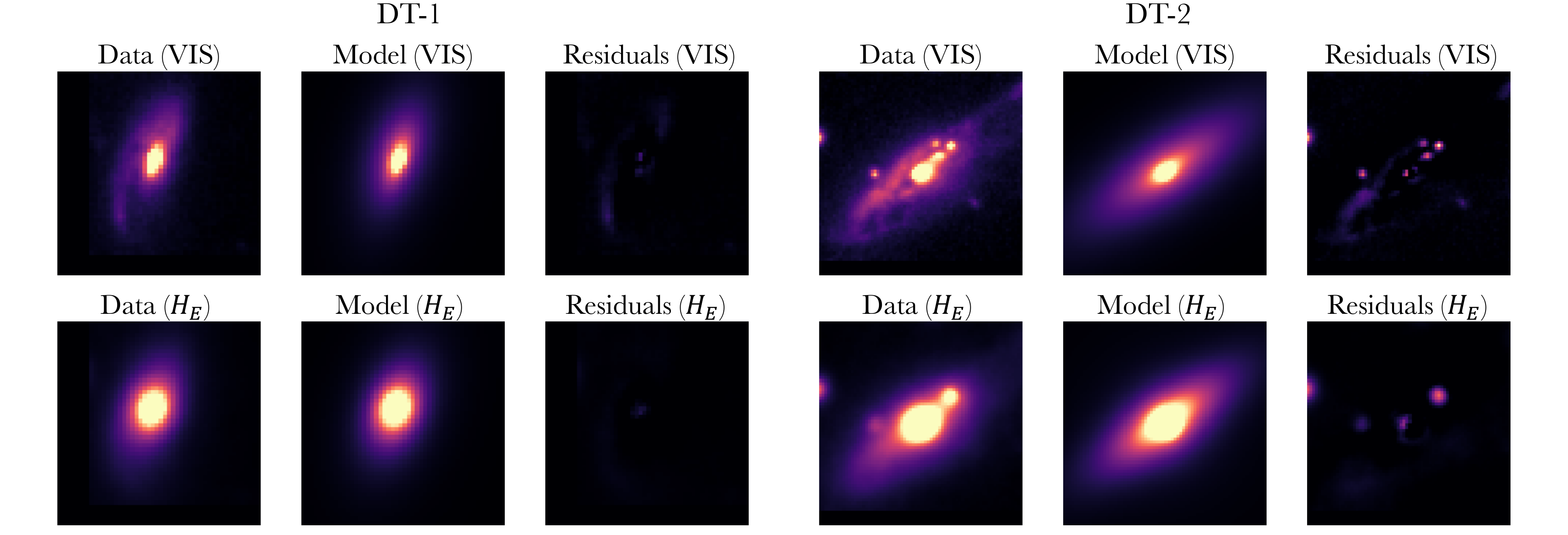}
    \caption{Morphological analysis of the two disky titans. For each galaxy, we report the data (in the \IE and \HE bands), the model with a double Sérsic profile (see Sect.\,\ref{sec:morph}), and the residuals. Each stamp has a size of $5\arcsec \times 5\arcsec$.}
    \label{fig:morph}
\end{figure*}

\subsection{SED fitting}
\label{sec:sed_fitting}

To supplement the physical properties estimated with the \texttt{nnpz} algorithm implemented in the \Euclid pipeline, we performed a traditional SED-fitting with \texttt{CIGALE} \citep{Boquien_19}. The chosen setup includes the single stellar population models by \citet{Bruzual_03}, combined through a delayed exponentially declining star formation history. Dust attenuation was taken into account following the model by \citet{Charlot_00}, while nebular emission was included using the set of \texttt{Cloudy} models \citep{Ferland_98} realised by \citet{Inoue_11}. The photometry given as input to \texttt{CIGALE} is that described in Sect.\,\ref{sec:euclid} and provided by the \Euclid MER pipeline\footnote{For consistency with \texttt{nnpz}, we used the photometry extracted in fixed circular apertures with a radius equal to twice the PSF full width at half maximum in the $u$ band with a wavelength-dependent aperture correction to account for the missing flux.} for the two galaxies considered as a whole, with the addition of an IRAC data point at $\lambda=3.6 \,\mu$m for DT-2 (see Sect.\,\ref{sec:euclid}). For the bulge and disk components of the two objects, we employed the photometry obtained through the procedure outlined in Sect.\,\ref{sec:morph}. In this case, the uncertainties are the sum in quadrature of the photometric uncertainties and of those related to the modelling procedure. The results of our SED-fitting procedure are shown in Fig.\,\ref{fig:seds}.

Examining the results for the galaxies when considered in their integrity (Fig.\,\ref{fig:comparison} and Table\,\ref{tab:prop_gal}), we notice how the sources are located at the high-mass end of the MS (according to the parametrisation by \citealt{Q1-SP031}). The high precision in the stellar mass estimates for the two objects is driven by the photometric coverage of the rest-frame NIR wavelengths provided by the $\HE$ filter (and the first channel of IRAC for DT-2). At the same time, the SFR of the two sources has larger uncertainties, mostly due to the lack of photometric constraints on the rest-frame UV regime. An alternative estimate of the SFR for these sources is from the measured continuum fluxes at $\lambda\sim2\,{\rm mm}$ (Sect.\,\ref{sec:spec_gas}). We converted these quantities into infrared luminosities through the standard templates for star-forming galaxies by \citet{Schreiber_18}. These quantities were then converted into SFRs by using the standard relation by \citet{Kennicutt_12}. The inferred values are reported in Table\,\ref{tab:prop_gal} and are consistent -- within the estimated uncertainties -- with those derived from SED-fitting. We emphasize that these values for the SFR were derived under the assumption that no significant AGN contribution is visible in the UV or at longer wavelengths. The assumption of no AGN contribution is based on the lack of X-ray detections for DT-2 (Fig.\,\ref{fig:xrays}) and radio detections at $150\,\mathrm{MHz}$ in the LOFAR maps covering EDF-N \citep{Bondi_24,Bisigello_25b} for both galaxies. {Regarding the X-rays, we performed a simple test with the software \texttt{PIMMS}\footnote{\url{https://heasarc.gsfc.nasa.gov/cgi-bin/Tools/w3pimms/w3pimms.pl}} to compute the expected X-ray flux of an AGN in our data. Assuming a low-luminosity AGN with $L_{\rm x}\sim10^{42.5} \, {\rm erg \, s^{-1}}$ and standard values $n_{\rm H}=10^{22} \, {\rm cm}^{-2}$ (hydrogen column density) and $\Gamma=1.7$ (index of the power-law spectrum), we obtain a total count of about eight photons in the energy range $0.5\mbox{--}2\,\mathrm{keV}$, below the detection limit of our observations. Therefore, a low-luminosity AGN cannot be excluded in DT-2 (and similarly in DT-1, due to the lack of X-ray coverage). However, we note that the presence of nuclear activity in our galaxies would not affect the two main quantities involved in our analysis: the SFR and the gas mass. The first could be affected if the rest-UV emission had a contribution from the accretion disk of the AGN, but the availability of a second tracer in the mm allows us to reject such a possibility.}

From the differential analysis of the two morphological components, we see that the star formation activity of the two galaxies is almost completely concentrated in the stellar disk, which contains $98\%$ and $70\%$ of the total SFR for the two galaxies, respectively. Similarly, the stellar mass is mostly confined to the bulges, accounting for $70\%$ and $84\%$ of the total stellar mass.

Moreover, additional information on the stellar populations hosted in our targets can be derived from the best-fitting SEDs. At the redshifts of our objects ($z\sim0.75$), the observed wavelength of the Balmer break falls between the $r$ and $i$ filters, making the $i-r$ colour a good estimator of the strength of the break. For this reason, we expect to have good constraints on the age of the main stellar population within the two galaxies from the SED-fitting, even in the absence of rest-optical spectroscopy in the region of the break. For both galaxies, the SED-fitting procedure shows that the bulge is significantly older than the disk, with stellar ages of the order of about $5\,{\rm Gyr}$, while the same quantities for the disks are of the order of $1\,{\rm Gyr}$.

Finally, the same constraints on the continuum emission at $\lambda\sim2 \, {\rm mm}$ can be used to measure the gas mass within our targets. Given that our observations sample the rest-frame $\lambda\sim1 \,{\rm mm}$ wavelength (on the Rayleigh--Jeans tail of the thermal dust emission), we can employ the observed fluxes to estimate the gas mass. This conversion is carried out through the same templates by \citet{Schreiber_18}, yielding dust masses of the order of $M_{\rm Dust}\sim10^{8-8.3} \, M_\odot$ (see Table \ref{tab:prop_gal}). These quantities can then be converted into dust-to-stellar mass ratios of $\log_{10}(M_{\rm Dust,1}/M_{\ast,1})=-3.3\pm0.4$ and $\log_{10}(M_{\rm Dust,2}/M_{\ast,2})=-3.0\pm0.3$ for the two targets, respectively. These values are slightly lower than what is commonly observed for MS galaxies. For instance, \citet{Donevski_20} reports an observed value of $\log_{10}(M_{\rm Dust,1}/M_{\ast,1})=-2.4\pm0.2$ for galaxies with analogous stellar masses and in the same redshift range 
(see analogous estimations in e.g. \citealt{Bethermin_15}).

\begin{figure*}
    \centering
    \includegraphics[width=0.48\linewidth]{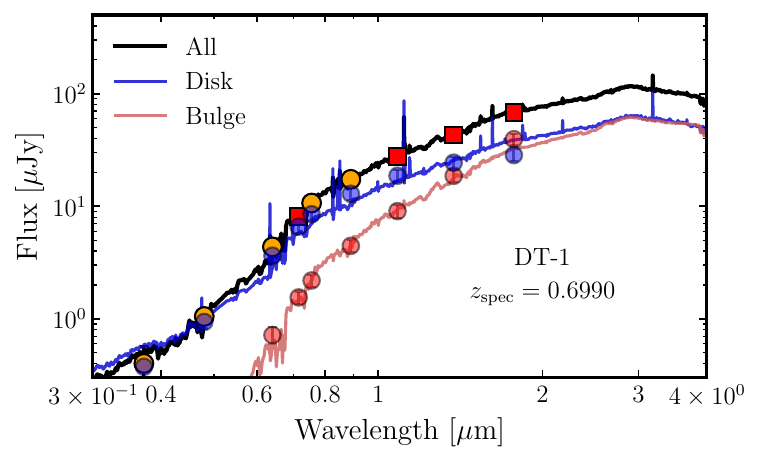} \ \includegraphics[width=0.48\linewidth]{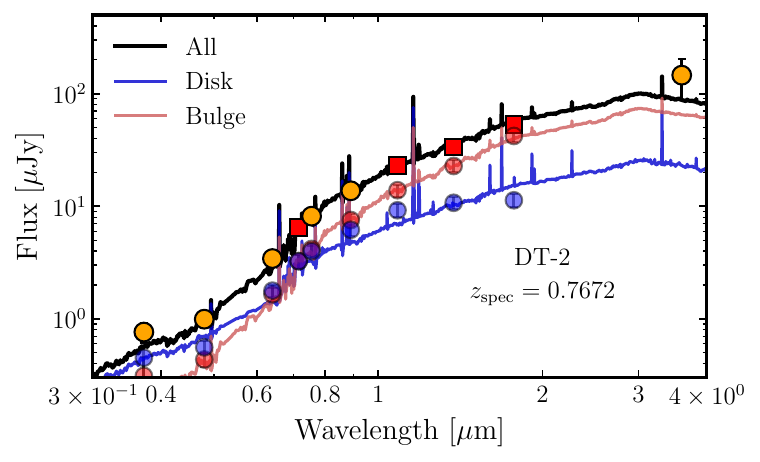}
    \caption{SED fitting performed with \texttt{CIGALE} on the two disky titans. Photometric data from \Euclid are represented as red squares, while additional photometry from ground-based observatories and IRAC is shown as orange points. The two coloured SEDs indicate the best-fitting SEDs for the bulge and disk components, with their respective photometry {(shown as shaded points of the corresponding colour)} derived as discussed in Sect.\,\ref{sec:morph}. For all SED fits, the redshift is fixed to the spectroscopic value reported in Sect.\,\ref{sec:spec_gas}.}
    \label{fig:seds}
\end{figure*}

\section{Characterisation of the groups}
\label{sec:prop_groups}

\begin{table}[]
    \centering
    \caption{Main properties estimated for the two groups analysed in this paper.}
    \label{tab:prop_groups}
    \begin{tabular}{lcc}
    & Group 1 & Group 2 \\
    \noalign{\vskip 1pt}
    \hline
    \noalign{\vskip 1pt}
    $N_{\rm gal}$ ($M_\ast>M_{\rm lim}$) & 10 & 10 \\
    $\log_{10}(M_\ast/M_\odot)$ (Vis.)  & $12.01\pm0.06$ & $11.94\pm0.04$ \\ 
    $\log_{10}(M_\ast/M_\odot)$ (Ext.)  & $12.05\pm0.06$ & $11.98\pm0.04$ \\ 
    \noalign{\vskip 1pt}
    \hline
    \noalign{\vskip 1pt}
    $r_{3, \rm ang}$ [$\arcsec$] & 49.3 & 31.6 \\ 
    $r_{3,\rm phys}$ [kpc] & 362 & 240 \\ 
    \noalign{\vskip 1pt}
    \hline
    \noalign{\vskip 1pt}
    $\log_{10}(M_{\rm h}/M_\odot)$ (opt.) & $13.8\pm0.4$ & $13.7\pm0.4$ \\
    $\log_{10}(M_{\rm h}/M_\odot)$ (X-rays) & \dots & $13.8\pm0.2$\\
    \noalign{\vskip 1pt}
    \hline
    \end{tabular}
    \tablefoot{The second row lists the cumulative stellar mass of the objects detected in the DR1 images of the two groups, while the third row lists the total stellar mass obtained by statistically accounting for less massive galaxies. The radii of the two groups correspond to the lengths at which the local density exceeds the median value at their redshifts by $3\sigma$. The two halo masses are computed, respectively, starting from the stellar mass of the group and from the X-ray luminosity (see Sect.\,\ref{sec:halo_mass}).}

\end{table}

\subsection{Estimation of the density field and group membership}
\label{sec:density}

The density field of EDF-N was estimated following the same procedure described in \citet{Q1-SP069} and is briefly summarised here. For each galaxy, we relied on the $\Sigma_N$ estimator, defined as
\begin{equation}
    \Sigma_N=\frac{N+1}{\pi R_N^2},
\end{equation}
where $R_N$ is the distance from the considered galaxy to the $N$-th closest neighbour. For consistency with \citet{Q1-SP069}, we employed the $\Sigma_5$ estimator applied to galaxies more massive than $10^{10.3}\,M_\odot$ and then interpolated to lower-mass galaxies. This choice allows us to focus on brighter objects with a more accurate photometric redshift estimation and accurately tracing the overall density field. Since our analysis mostly relies on photometric redshifts (which do not allow 3D modelling of the density field), we adopted on a tomographic approach. Given our interest in two specific over-densities, we employed thin redshift slices, defined as
\begin{equation}
    |z-z_{\rm DT}|<0.03 \, (1+z_{\rm DT}),
\label{eq:redshift}
\end{equation}
where $z_{\rm DT}$ is the spectroscopic redshift of the disky titan considered. A factor of 0.03 accounts for the median accuracy of the photometric redshifts computed by the \Euclid pipeline \citep{Q1-SP031}. We consider the density contrast parameter, defined as
\begin{equation}
    \log_{10}(1+\delta)=\log_{10}\left(1+\frac{\Sigma_N - \bar{\Sigma}_N}{\bar{\Sigma}_N}\right),
\end{equation}
where $\bar{\Sigma}_N$ is the median density field of a given redshift slice. This parameter follows a Gaussian distribution, as already shown, for instance, in \citet{Q1-SP069}. As expected from the selection scheme (Sect.\,\ref{sec:sample_selection}), the two targets belong to strong over-densities, with $\log_{10}(1+\delta)$ of 1.4 and 1.0, corresponding to $6\sigma$ and $5\sigma$ above the median density at their redshifts. Quite remarkably, both galaxies are not found at the projected centre of the two over-densities (as indicated by the density curves shown in Fig.\,\ref{fig:groups} for DT-1 and by the detected X-ray emission for DT-2; see Sect.\,\ref{sec:xrays}), despite being the brightest sources in their groups.

\subsection{Estimation of the halo mass}
\label{sec:halo_mass}

\begin{figure}
    \centering
    \includegraphics[width=\columnwidth]{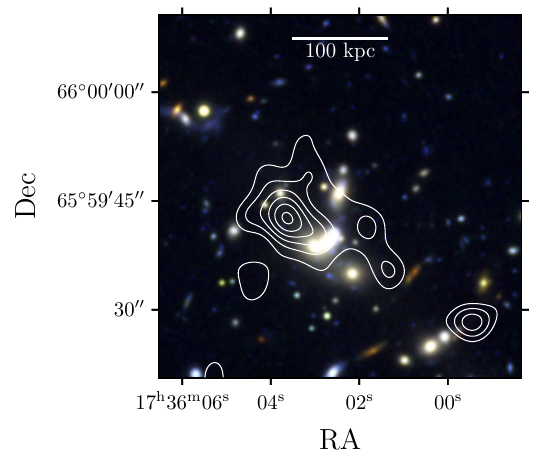}
    \caption{Diffuse X-ray emission of the massive group hosting DT-2, as observed by XMM-Newton. The contours show the X-ray emission in the soft band ($0.5\mbox{--}2\,\mathrm{keV}$) in units of $\sigma$ above the median, starting from $5\sigma$ and with $2\sigma$ steps. The X-ray spectrum is consistent with {a collisionally ionised diffuse gas} with $T\sim3\times10^7\,{\rm K}$ and a total $L_{\rm X}\sim2\times10^{43}\,{\rm erg\,s^{-1}}$.} 
    \label{fig:xrays}
\end{figure}

The halo masses of the two structures analysed in this paper were computed assuming virialised halos, following a slightly modified version of the procedure described in \citet{Daddi_21} and \citet{Sillassen_22}. First, we summed all the stellar masses of the galaxies possibly belonging to the two groups. To do this, we considered all objects within $r_3$ (i.e. the region where the logarithmic local density field is above $3\sigma$ from its median value) as candidate members, with a redshift consistent with the $z_{\rm spec}$ of the respective DT, according to Eq.\,(\ref{eq:redshift}). Given the mass completeness estimated for our data (see Sect.\,\ref{sec:euclid}), we corrected the total stellar mass by integrating the stellar mass function at $z\sim0.7$ by \citet{Shuntov_25} down to $10^7 \, M_\odot$ (corresponding to a small correcting factor of 0.91). The total stellar masses of the two groups were then converted into halo masses of $\sim10^{13.8} \, M_\odot$ using the conversion between stellar mass and halo mass by \citet{Van_der_burg_14} calibrated for clusters at $z\sim1$.\footnote{Although the redshift range analysed in the authors' study is a bit higher than that of our targets, we notice that their results are consistent with those obtained by \citet{Lin_12} for clusters at $z<0.6$. This ensures that the relation does not significantly change at low $z$.} The derived halo masses -- together with the other estimated properties of the two groups -- are reported in Table \ref{tab:prop_groups}. 

For the second group hosting DT-2, the availability of X-ray data (see Sect.\,\ref{sec:xrays}) provides better constraints on the halo mass. First, the detection of diffuse X-ray emission is likely due to the presence of hot gas, a strong indicator of the expected virialisation of the group halo (see e.g. \citealt{Sarazin_68} and references therein). Second, the integrated X-ray luminosity can be linked to the halo mass using standard scaling relations (see e.g. \citealt{Lovisari_21} and references therein). To do so, for each XMM EPIC camera, we extracted the X-ray spectrum of the emission from our over-density in a circular aperture encompassing the emission visible in the stacked image (Fig.\,\ref{fig:xrays}). The background spectra were extracted from fixed apertures of the same size located in an empty region of the nearby sky. We then performed spectral modelling with \texttt{XSPEC} \citep{Arnaud_96}. We employed an Astrophysical Plasma Emission Code (APEC; \citealt{Smith_01}) model describing a collisionally ionised diffuse gas. The observed spectrum is consistent with an intra-cluster medium at a temperature of $T=3^{+2}_{-1}\times10^7\, {\rm K}$ (see Appendix\,\ref{app:xrays}). The best-fitting spectrum yields a total integrated luminosity of $L_{\rm 0.1\mbox{--}2.4 \, keV}=(2.2\pm0.3) \times 10^{43} \, {\rm erg \, s^{-1}}$ in the energy range $0.1\mbox{--}2.4\,\mathrm{keV}$, assuming the group is located at the same spectroscopic redshift as the most massive galaxy. Through the scaling relation by \citet{Lovisari_21}, linking the X-ray luminosity to the halo mass, we estimate the latter as $\log_{10}(M_{\rm h}/M_\odot)=13.8\pm0.2$. This result is consistent -- within the estimated uncertainties -- with the values inferred from the optical data.

\section{Discussion}
\label{sec:discussion}

The evidence presented in the previous sections of this paper pictures the two disky titans as massive ($M_\ast>10^{11} \, M_\odot$), star-forming galaxies (i.e. in the MS) located in strong over-densities (above $5\sigma$ from the median density field). These objects are quite rare: in over $20\,\mathrm{deg}^2$ of EDF-N, only four objects located in the redshift range $0.5<z<0.75$ show similar characteristics \citep{Q1-SP069}. In this section, we discuss possible scenarios to explain the properties of these objects.

A first point to address concerns the molecular gas mass observed in our targets. These values are consistent with those commonly found in MS galaxies, as discussed in Sect.\,\ref{sec:spec_gas}. This is quite uncommon for BGGs of massive groups, which are commonly gas-poor (e.g. \citealt{Castignani_20,Castignani_22,Castignani_23}). In the standard cold accretion paradigm, the star formation activity of MS galaxies is sustained over long timescales by the accretion of cold gas from the surrounding environment. The high halo masses of massive groups such as those hosting our targets make this explanation unfeasible. Several studies have already indicated how the cold fraction of the baryonic accretion rate (BAR) -- that is, the component available for star formation without requiring a long cooling time -- typically decreases with increasing halo mass as an effect of virial shock heating (e.g. \citealt{Dekel_06,Correa_18}). To analyse quantitively how this phenomenon impacts our galaxies, we computed the BAR through the parametrisation using \citet{Goerdt_10}
\begin{equation}
    {\rm BAR}\sim137\left(\frac{M_{\rm h}}{10^{12} \, M_\odot}\right)^{1.15}\left(\frac{1+z}{4}\right)^{2.25} \, M_\odot\,{\rm yr^{-1}} \ .
\end{equation}

Then, we used the parametrisation by \citet{Correa_18} to estimate the cold fraction
\begin{equation}
    f_{\rm cold}= 1 - \frac{1}{1+[M_{\rm h}/M_0(z)]^{a(z)}} \ ,
\end{equation}
with the redshift-dependent expressions for $a(z)$ and $M_0$ taken from \citet{Correa_18} in the $0<z<2$ case. The computation tells us that, in this framework, both galaxies should be accreting a negligible amount of cold gas from the surrounding environment (less than $1\, M_\odot\,{\rm yr}^{-1}$), which is clearly insufficient to sustain a SFR of about $20\,M_\odot\,{\rm yr}^{-1}$ (see Table\,\ref{tab:prop_gal}) over long timescales. This holds even with unphysically high SFEs -- significantly above the 20\% efficiency commonly observed at low redshifts (e.g. \citealt{Shuntov_25}) and the potentially higher values inferred in high-\textit{z} studies (e.g. \citealt{Xiao_24,Gentile_24b}).

Once excluding the cold accretion scenario, an additional possibility for explaining the molecular gas in our targets could be the so-called cooling flows (see e.g. the review by \citealt{Fabian_94} and references therein). These flows of cold gas from the outskirts of massive groups and clusters to the centre of their potential well are thought to be caused by the radiative cooling of the intra-cluster medium and the consequent low pressure originating at the centre of the cluster. Several studies have suggested that these flows could reach the BCG and sustain its star formation (e.g. \citealt{Donahue_15,McDonald_18}). However, our galaxies are not located at the centre of their over-densities, as visible from the density curves in Fig.\,\ref{fig:groups} and from the X-ray data shown in Fig.\,\ref{fig:xrays}, {with a projected offset of the order of about 90 proper kpc. Even though examples of offset cooling flows have been reported in the current literature (e.g. as an effect of sloshing; \citealt{Pasini_19}), the projected distances are significantly lower than those observed in our disky titans (of the order of 6 kpc; \citealt{Pasini_19}). For this reason, this explanation is disfavoured (even though not completely excluded) for our sources.}\footnote{ {We note, however, that recent studies by \citet{Peiris_26} and \citet{Pontzen_26} theorised a modified mechanism where, above a certain value of halo mass ($M_h\sim10^{13.5-14.0} \ M_\odot$), the entropy of the circumgalactic medium reaches a level that cannot be sustained by AGN feedback. This opens the possibility of subsequent gas cooling in the most massive galaxy of an over-dense environment, potentially sustaining star formation.}} A third possibility for the origin of the molecular gas includes accretion from other galaxies. This scenario is discussed in further detail in Sect.\,\ref{sec:rejuvenation}.

A final point to address concerns the SFRs and efficiencies observed in our galaxies. According to the mechanism proposed by \citet{Martig_09}, the presence of a massive bulge in our galaxies should help stabilise the stellar disk, preventing star formation activity. Several studies and simulations have demonstrated how this morphological quenching cannot completely quench galaxies \citep[e.g.][]{Saintonge_12,Davis_14,Gobat_18,Eales_20}; nevertheless, its effect should still be visible as a reduction of the SFE (up to a factor of $3\mbox{--}10$; e.g. \citealt{Gobat_18}) with respect to the Schmidt--Kennicutt relation for star-forming galaxies (Fig.\,\ref{fig:comparison}).

\subsection{Two rejuvenating massive galaxies?}
\label{sec:rejuvenation}

A possible model explaining all the observables collected for the two disky titans involves the process of rejuvenation. This term is commonly defined as the phenomenon in which a quiescent galaxy restarts its star formation activity by accreting cold gas either from the surrounding environment or from another source. { Following the points discussed in the previous sections, scenarios where rejuvenation is driven by cold gas accretion from the cosmic web or through cooling flows are disfavoured, leaving us the option that the cold gas comes from another galaxy during a merging episode.} Several simulations \citep{Robertson_06,Governato_09,Hopkins_09} and some observational studies (e.g. \citealt{Belli_17,Mancini_19,Wang_25,Zhuang_25}) have shown that a wet merger can enable a galaxy to re-build a stellar disk and obtain levels of star formation high enough to place it in the green valley or on the MS.

Specifically, our observations can be inscribed into a scenario where a `classic' massive ($M_\ast\sim10^{11} \ M_\odot$), passive, and spheroidal BGG merges with a gas-rich satellite resulting in the accretion of cold gas onto the more massive galaxy. This process produces a temporary increase in the molecular gas fraction of the BGG and an increase in the SFE driven by the gas compression during the accretion phase (as shown in several simulation-based studies describing analogous enhancements in galaxy merging; e.g. \citealt{DiMatteo_07,Moreno_21,Schechter_25}). The following stabilisation of the disk driven by the presence of the massive bulge \citep{Martig_09,Gobat_18} thus produces a reduction in the observed SFE and the formation of a standard passive disk (commonly observed at low redshifts; e.g. \citealt{Masters_10}). The eventual transition of these objects into classic spheroidal BGGs can then be driven by external mechanisms such as dry mergers with other satellite galaxies. A recent study by \citet{Zhuang_25} presented direct evidence of such a mechanism by observing the accretion of a small, gas-rich satellite onto a more massive, quiescent galaxy using high-resolution {\it James Webb} Space Telescope (JWST) observations. 

In our case, such a scenario is consistent with the evidence presented in previous sections of this paper. Specifically:
\begin{enumerate}
    \item Both galaxies present a passive bulge, with an SFR lower than expected for MS galaxies with the same mass (Sect.\,\ref{sec:sed_fitting} and Fig.\,\ref{fig:comparison}). The Sérsic index of this component is high ($n_{\rm bulge}\sim6$), consistent with a structure supported by velocity dispersion, as expected for massive spheroidal galaxies at low redshifts \citep{Kormendy_04,Fisher_08}. Moreover, the strength of the Balmer break (measured from photometric data and thus quite uncertain) potentially indicates that the stellar population inside the bulge is significantly older than that in the disk (Sect.\,\ref{sec:sed_fitting}).

    \item For both galaxies, the molecular gas mass is of the order of $10^{10\mbox{--}10.5} \, M_\odot$. Given the normal gas fraction expected from MS galaxies \citep[e.g.][]{Tacconi_20}, these gas masses are consistent with those expected from less massive star-forming galaxies up to $10^{9.5-10} \, M_\odot$. At the same time, the low dust-to-stellar mass ratios compared to analogous MS galaxies indicates a possible ex situ origin for a part of the stellar mass.

    \item Both galaxies present a slightly distorted morphology. DT-1 presents a tidal feature in the external part of the disk, while DT-2 shows a small asymmetry in the spiral arms of the disk (Fig.\,\ref{fig:morph}). Both features can be interpreted as the result of a close encounter with another object.
\end{enumerate}

The two targets would therefore represent two stages of the same evolutionary scenario: DT-1 would have been observed shortly after the close encounter with the satellite, easily explaining the more disturbed morphology and the higher SFE. DT-2, instead, would represent a later stage where the SFE returns to standard values and the morphology typically stabilises. In addition, the morphological analysis of DT-1 shows a NIR-bright clump (see the residual panel in Fig.\,\ref{fig:morph}) that could be interpreted as a residual of the bulge of the accreted galaxy, supporting the merger-induced rejuvenation scenario. This possibility, however, must be confirmed with follow-up spectroscopic observations. Alternatively, another galaxy with a redshift consistent with the group and located at a projected distance of about 200 kpc south of DT-1 also shows a significant tidal feature. The galaxy's distance would allow for a close encounter with DT-1, given the velocity dispersion expected for a massive group with a halo mass of $M_{\rm h}\sim10^{13.8} \, M_\odot$ (applying the virial theorem, about $750 \, \kms$) and the age of the disk in DT-1 (Sect.\,\ref{sec:sed_fitting}).

Such a scenario also fits the current consensus on the evolutionary path that produces the formation of the most massive BGG, with most of the stellar mass assembled at high redshifts ($z\sim3\mbox{--}5$; e.g. \citealt{DeLucia_07}) and a smaller fraction assembled at later time (after $z\sim1$; e.g. \citealt{Collins_09,Stott_11}) through the accretion of satellite galaxies. Finally, the proposed scenario easily explains the observed rarity in this type of source by relying on a non-secular process such as a galaxy merger. 

\subsection{Comparison with simulation}

\begin{figure}
    \centering
    \includegraphics[width=\linewidth,trim={0 0.7cm 0 0},clip]{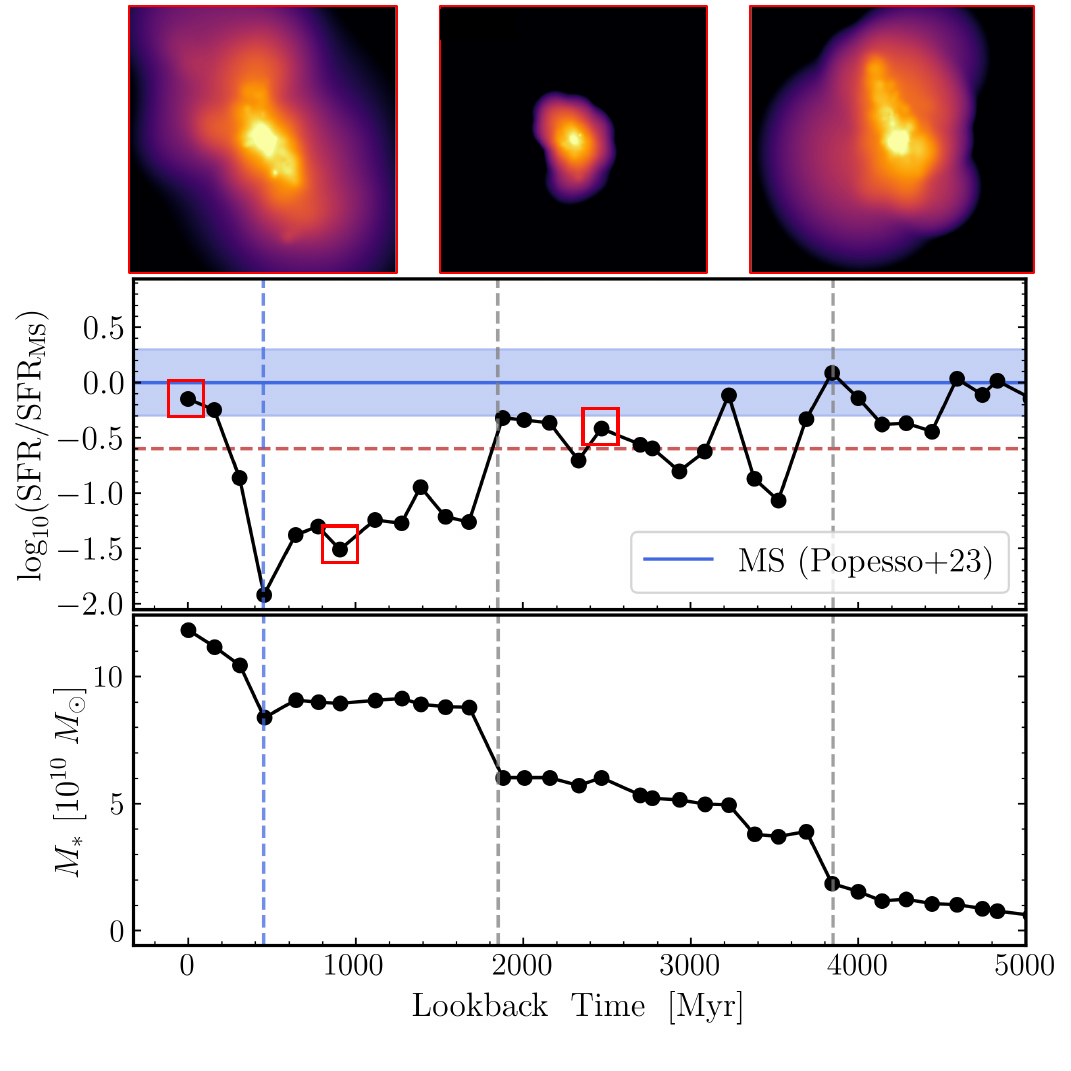}
    \caption{Example of an analogue of the disky titans in the TNG300 simulation. \emph{Bottom panel:} Growth in stellar mass over the last $5\,\mathrm{Gyr}$. The vertical dashed lines indicate the past mergers, with the last episode causing rejuvenation highlighted in blue. \emph{Middle panel:} Star formation history, parametrised as the distance from the MS by \citet{Popesso_23}, where the MS (and its intrinsic scatter) is indicated in blue and the red sequence (below $2\sigma$ from the relation) is indicated as a dashed red line. \emph{Top panel:} Density maps of the stellar component in the TNG analogue at key cosmic epochs, the corresponding instants in the star formation history are highlighted by the red boxes. The cutouts have a constant size of $500\,\mathrm{kpc}$.}
    \label{fig:TNG}
\end{figure}

{To verify the plausibility of our scenario, we searched for analogues of our disky titans in the IllustrisTNG simulation suite \citep{Nelson_19}. Given the requirement for our galaxies to be located in over-dense environments, we analysed the TNG300 simulation, which includes a cosmological volume of about $300\,\mathrm{Mpc}^3$ at a (baryonic) resolution of $m_{\rm b}=1.1 \times10^{7} \, M_\odot \, h^{-1}$. This resolution is ideal for sampling a large variety of environments. The selection criteria mimic those employed in Sect.\,\ref{sec:sample_selection}: we selected galaxies from the snapshot at $z=0.76$ with $M_\ast>10^{11} \, M_\odot$, within $0.3\,\mathrm{dex}$ from the MS, and located in over-densities. For the second criterion, we employed the parametrisation of the MS by \citet{Q1-SP031} and the SFRs averaged over $100\,\mathrm{Myr}$ computed by \citet{Donnari_19} and \citet{Pillepich_19} to better match the observational constraints. The last criterion was accounted for through a cut in the mass of the hosting dark matter ($M_{\rm h}>10^{13} \, M_\odot$), approximately corresponding to the expected halo mass of a $5\sigma$ over-density at $z\sim0.75$ \citep{Q1-SP069}. These criteria yield a total sample of 238 galaxies, with a median halo mass of $M_{\rm h}\sim10^{13.5} \, M_\odot$ and representing about $10\%$ of the galaxies with similar halos, in good agreement with the findings presented in the previous sections. For these analogues, we analysed the evolution with time of the stellar mass, SFR, halo mass, and gas mass in the last $5\,\mathrm{Gyr}$ prior to the observations (roughly corresponding to $z\sim3$). We find that in nearly half of the cases, these sources spent a significant amount of time (about $2\mbox{--}3\, \mathrm{Gyr}$) in the red sequence before experiencing an episode of rejuvenation and returning to the MS. In one quarter of the cases, the decrease in the specific SFR was not sufficient to bring the galaxies to the red sequence but only to the green valley (i.e. between one and two $\sigma$ below the MS). In the remaining cases, no significant deviation from the secular evolution in the MS is visible in the TNG data.

The first case clearly agrees with the scenario outlined in the previous sections. We report an example galaxy belonging to this class in Fig.\,\ref{fig:TNG}. We note how this source, after a first period of secular evolution as part of the MS (up to a look-back time of $4\, \mathrm{Gyr}$; $z\sim2$), undergoes a first merger episode and moves to the red sequence. In the next $2\,\mathrm{Gyr}$, the accretion of small satellites drives an increase in its stellar mass and the oscillation between the red sequence and the green valley, up to a look-back time of $2\,\mathrm{Gyr}$ ($z\sim1.3$). In this phase, the sequence of small merging episodes produces a disturbed morphology (rightmost cutout). Then, a second merger episode causes another transition to the red sequence, for a long period of quiescence lasting about $2\,\mathrm{Gyr}$ in which the morphology relaxes and the size becomes more compact (central cutout). In this phase, the galaxy appears as a typical BGGs at these redshifts. Finally, about $500\,\mathrm{Myr}$ before the observation, a last merging episode causes the rejuvenation of the galaxy and its return to the MS. The accretion of the gas from the satellite and the new star formation activity take place in the outskirts of the galaxy, producing a more extended morphology (leftmost cutout), analogous to that observed in our two sources.}

\section{Summary}
\label{sec:summary}

In this paper, we presented the discovery of two disky titans in the first data from the \Euclid satellite. These galaxies are defined as massive ($M_\ast>10^{11} \, M_\odot$) and star-forming (${\rm SFR}\sim20 \, M_\odot \, {\rm yr}^{-1}$) systems located in strong over-densities (exceeding $5 \sigma$ from the median density) at $z\sim0.75$ and showing a visible stellar disk. These sources are quite rare, with only four candidates observed in over $20\,\mathrm{deg}^2$ covered by \Euclid in EDF-N.

Our observational results can be summarised as follows:
\begin{itemize}

    \item Both galaxies are the brightest sources in two massive group-size structures with $M_{\rm h}\sim10^{13.8}\,M_{\odot}$. According to the predictions for virial-shock heating, both sources are expected to accrete only a negligible amount of cold gas from the surrounding environment, insufficient to sustain significant star formation over long time-scales.
    
    \item Despite this, the two galaxies host significant gas reservoirs ($M_{\rm H_2}\sim10^{10.3} \, M_\odot$). Moreover, both sources show an SFE and a depletion time consistent with that commonly expected for MS galaxies according to the Schmidt--Kennicutt relation, or a slightly higher SFE in the case of DT-1.

    \item Both galaxies present a massive, passive bulge at their centre, with stellar ages higher than those observed in the disk. In both galaxies, the bulk of the stellar mass is located in the bulge, while star formation activity is confined within the stellar disk.
\end{itemize}

We interpret these observations as evidence that our disky titans are the product of an episode of merging-induced rejuvenation, in which the most massive galaxy of the groups accretes cold gas from a surrounding member of the structure and restarts its star formation activity. This scenario easily explains the observed properties of the two sources and justifies the observed MS-like SFE as a composite effect of a merger-induced starburst and morphological quenching. Such a scenario {is supported by the identification in the TNG300 simulation of analogous systems with evidence of past merger activity linked to rejuvenation. However, the proposed scenario} requires further observations to be completely confirmed. Spatially resolved optical spectroscopy (e.g. through the integral field unit equipped on the JWST or with VLT-MUSE) would allow us to measure the stellar ages of the bulge and disk of our galaxies, confirming the initial picture inferred from broadband photometry alone. Similarly, deeper mm observations would allow us to better constrain the continuum emission of our sources, giving us a better handle on the obscured star formation activity. At the same time, higher-resolution observations of the (sub)mm continuum and of the CO(2-1) line would allow us to determine whether the gas and star formation are actually located in the outer disk or whether a component is located in the central bulge. Such data would also enable dynamical modelling of the gas component, giving us an additional probe of the dynamical mass independent of the stellar mass derived from SED fitting. Both the observations listed so far (resolved spectroscopy and deeper NOEMA observations) could also allow us to detect the molecular gas undergoing accretion on our main targets, either by observing a low-$J$ CO transition or by detecting the Ly$\alpha$ line (e.g. \citealt{Daddi_22b,Guo_25}). Finally, the identification of the other members of the group still relies only on photometric redshift. A more intensive spectroscopic coverage of the environment surrounding our galaxies would allow us to reduce the uncertainty on the estimation of the halo mass and confirm the inferred lack of cold accretion from the surrounding environment.

In summary, our study showcases the ability of \Euclid to find rare sources thanks to the unprecedented statistics offered by its surveys. Moreover, the synergy between \Euclid and other facilities observing the Universe at other wavelengths gives us the ability to test our galaxy evolution models in new and unexplored regions of the parameter space.

\begin{acknowledgements}
This work is based on observations carried out under project number S25BT with the IRAM NOEMA Interferometer. IRAM is supported by INSU/CNRS (France), MPG (Germany) and IGN (Spain).
FaGe, EmDa, AnEn, GaDL, ChD'E, LuPo acknowledge support from the ELSA project. "ELSA: Euclid Legacy Science Advanced analysis tools" (Grant Agreement no. 101135203) is funded by the European Union. Views and opinions expressed are however those of the author(s) only and do not necessarily reflect those of the European Union or Innovate UK. Neither the European Union nor the granting authority can be held responsible for them. UK participation is funded through the UK HORIZON guarantee scheme under Innovate UK grant 10093177.
VaSa acknowledges support from France 2030 through the project named Académie Spatiale d’Île-de-France (https://academiespatiale.fr/) managed by the National Research Agency under bearing the reference ANR-23-CMAS-0041
\AckEC
\AckDRone

Based on data from UNIONS, a scientific collaboration using three Hawaii-based telescopes: CFHT, Pan-STARRS, and Subaru \url{www.skysurvey.cc}\,. Based on data from the Dark Energy Camera (DECam) on the Blanco 4-m Telescope at CTIO in Chile \url{https://www.darkenergysurvey.org}\,.
The IllustrisTNG simulations were undertaken with compute time awarded by the Gauss Centre for Supercomputing (GCS) under GCS Large-Scale Projects GCS-ILLU and GCS-DWAR on the GCS share of the supercomputer Hazel Hen at the High Performance Computing Center Stuttgart (HLRS), as well as on the machines of the Max Planck Computing and Data Facility (MPCDF) in Garching, Germany.

\end{acknowledgements}

\bibliography{biblio}

\begin{appendix}

\section{Additional figures}
\label{app:xrays}
Figure\,\ref{fig:rgb} shows the two additional disky titans observed in EDF-N but not analysed in detail in the paper. Figure\,\ref{fig:fitting} reports the best-fitting spectrum for the XMM data of DT-2. 

\begin{figure}[h]
    \centering
    \includegraphics[width=0.4\linewidth]{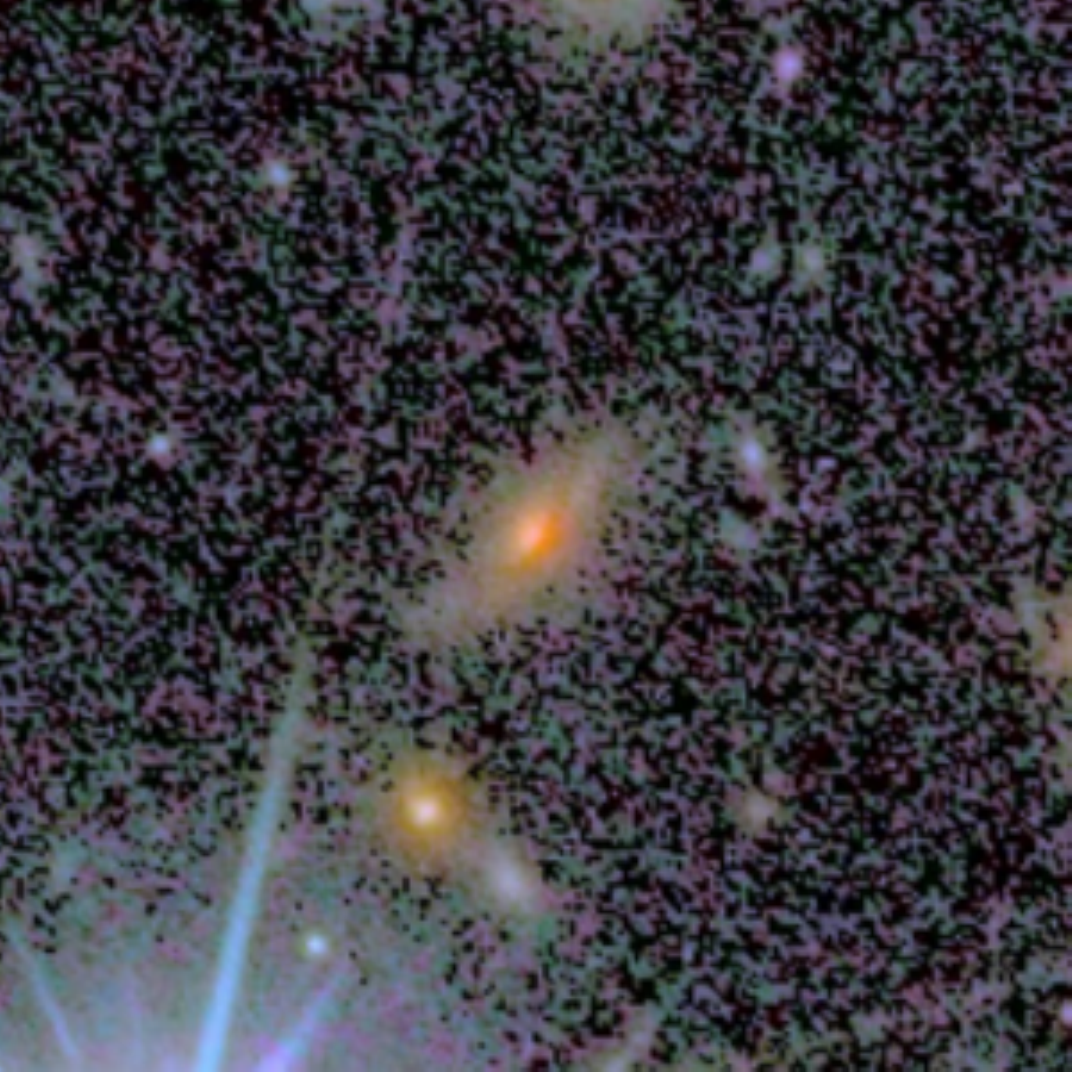} \, \includegraphics[width=0.4\linewidth]{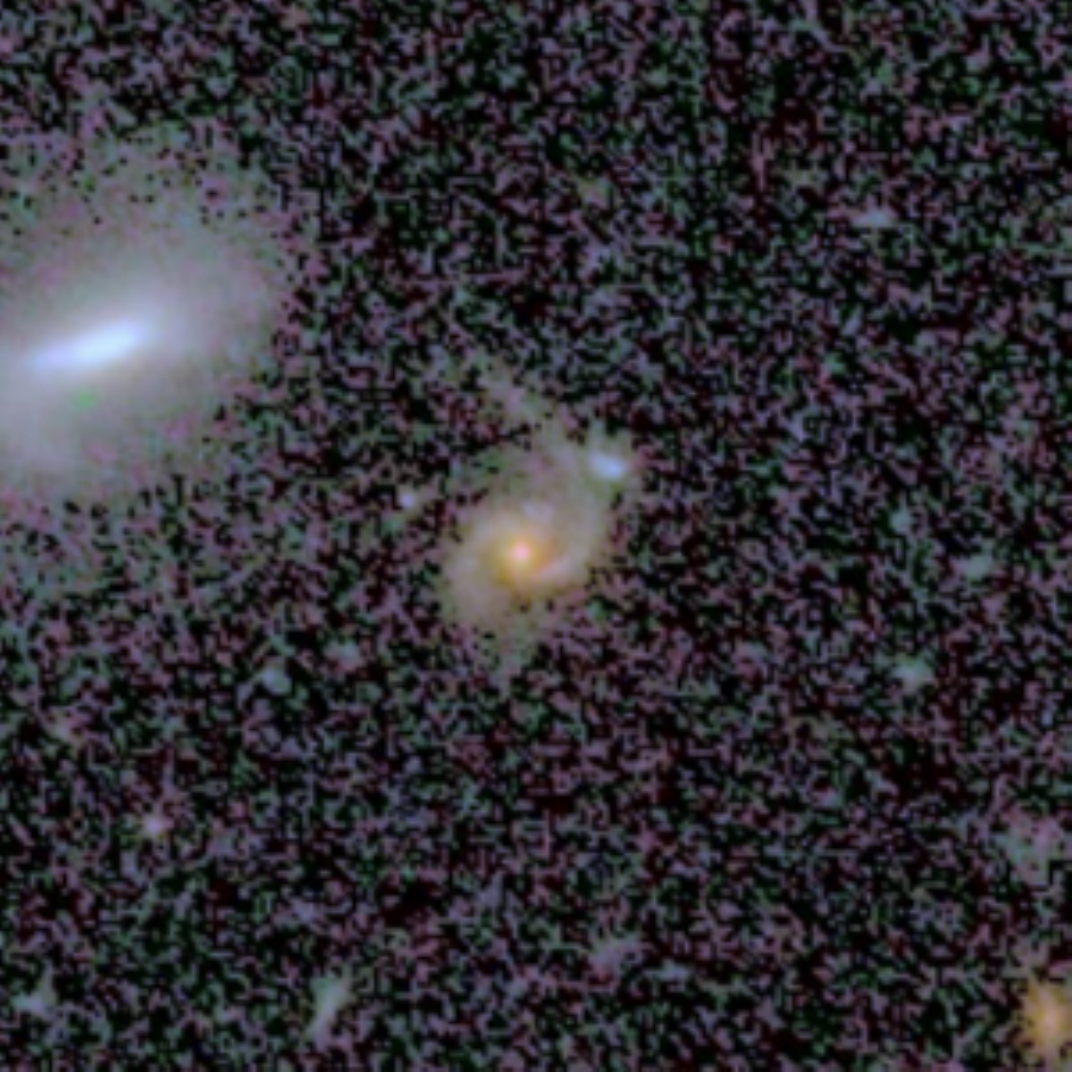}
    \caption{RGB images of the two additional disky titans observed in EDF-N but not analysed in detail in this study. The cutouts have a side of 20 arcsec and are generated following the same procedure as in Fig.\,\ref{fig:groups}.}
    \label{fig:rgb}
\end{figure}
    
\begin{figure}[h]
    \centering
    \includegraphics[width=\linewidth]{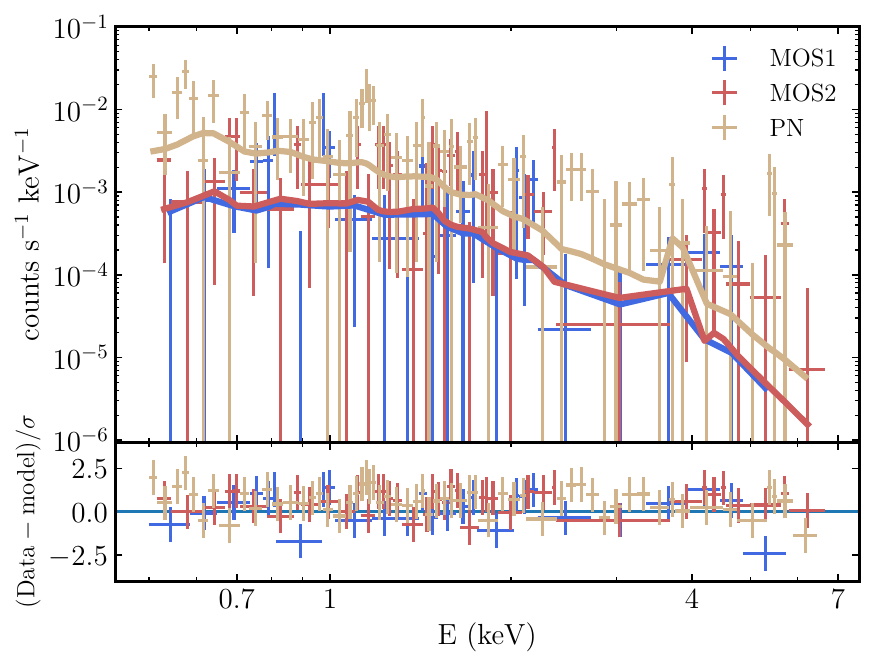}
    \caption{Best-fit spectrum for the X-ray data of DT-2. The coloured crosses report the measurements of the MOS-1, MOS-2, and NP instruments of XMM (blue, red, and gold, respectively), while the corresponding solid lines report the best-fitting model for each instrument: a collisionally ionised diffuse gas with $T\sim3\times10^7\,{\rm K}$ and a total $L_{\rm X}\sim2\times10^{43}\,{\rm erg\,s^{-1}}$. \emph{Bottom panel}: Normalised residuals.}
    \label{fig:fitting}
\end{figure}

\end{appendix}

\end{document}